\documentclass[fleqn,usenatbib]{mnras}

\usepackage{newtxtext,newtxmath}
\usepackage[T1]{fontenc}

\DeclareRobustCommand{\VAN}[3]{#2}
\let\VANthebibliography\thebibliography
\def\thebibliography{\DeclareRobustCommand{\VAN}[3]{##3}\VANthebibliography}

\usepackage{graphicx}	% Including figure files
\usepackage{amsmath}	% Advanced maths commands
\usepackage{amssymb}	% Extra maths symbols
\usepackage{longtable}
\usepackage{tablefootnote}
\title[MeerKAT Variable Sources]{Search and identification of transient and variable radio sources using MeerKAT observations: a case study on the MAXI J1820+070 field}

\author[A. Rowlinson et al.]{A. Rowlinson,$^{1,2}$\thanks{E-mail: b.a.rowlinson@uva.nl}
J. Meijn,$^{1}$ J. Bright,$^{3}$ A.J. van der Horst,$^{4}$ S. Chastain,$^{4}$ S. Fijma,$^{1}$ R. Fender,$^{5}$  \and I. Heywood,$^{5,6,7}$ R.A.M.J. Wijers,$^{1}$ P.A. Woudt,$^{8}$ A. Andersson,$^{5}$ G.R. Sivakoff,$^{9}$ E. Tremou,$^{10}$  \and L.N. Driessen,$^{11}$
\\
$^{1}$ Anton Pannekoek Institute, University of Amsterdam, Postbus 94249, 1090 GE Amsterdam, The Netherlands\\
$^{2}$ ASTRON, the Netherlands Institute for Radio Astronomy, Oude Hoogeveensedijk 4, 7991 PD, Dwingeloo, The Netherlands\\
$^{3}$ Department of Astronomy, University of California, Berkeley, CA 94720-3411, USA\\
$^{4}$ Department of Physics, The George Washington University, 725 21st Street NW, Washington, DC 20052, USA \\
$^{5}$ Astrophysics, Department of Physics, University of Oxford, Keble Road, Oxford OX1 3RH, UK\\
$^{6}$ Department of Physics and Electronics, Rhodes University, PO Box 94, Makhanda 6140, South Africa\\
$^{7}$ South African Radio Astronomy Observatory, 2 Fir Street, Observatory 7925, South Africa\\
$^{8}$ Department of Astronomy, University of Cape Town, Private Bag X3, Rondebosch 7701, South Africa\\
$^{9}$ Department of Physics, University of Alberta, CCIS 4-181, Edmonton, AB T6G 2E1, Canada\\
$^{10}$ National Radio Astronomy Observatory, P.O. Box O, Socorro, NM 87801, USA \\
$^{11}$ CSIRO, Space and Astronomy, PO Box 1130, Bentley, WA 6102, Australia \\
}

\date{Accepted XXX. Received YYY; in original form ZZZ}

\pubyear{2015}

\begin{document}
\label{firstpage}
\pagerange{\pageref{firstpage}--\pageref{lastpage}}
\maketitle

\begin{abstract}
Many transient and variable sources detected at multiple wavelengths are also observed to vary at radio frequencies. However, these samples are typically biased towards sources that are initially detected in wide-field optical, X-ray or gamma-ray surveys. Many sources that are insufficiently bright at higher frequencies are therefore missed, leading to potential gaps in our knowledge of these sources and missing populations that are not detectable in optical, X-rays or gamma-rays. Taking advantage of new state-of-the-art radio facilities that provide high quality wide-field images with fast survey speeds, we can now conduct unbiased surveys for transient and variable sources at radio frequencies. 
In this paper, we present an unbiased survey using observations obtained by MeerKAT, a mid-frequency ($\sim$GHz) radio array in South Africa's Karoo Desert. The observations used were obtained as part of a weekly monitoring campaign for X-ray binaries (XRBs) and we focus on the field of MAXI J1820+070. We develop methods to efficiently filter transient and variable candidates that can be directly applied to other datasets. In addition to MAXI J1820+070, we identify four likely active galactic nuclei, one source that could be a Galactic source (pulsar or quiescent X-ray binary) or an AGN, and one variable pulsar. No transient sources, defined as being undetected in deep images, were identified leading to a transient surface density of $<3.7\times10^{-2}$ deg$^{-2}$ at a sensitivity of 1 mJy on timescales of one week at 1.4 GHz.
\end{abstract}


%%%%%%%%%%%%%%%%%%%%%%%%%%%%%%%%%%%%%%%%%%%%%%%%%%

%%%%%%%%%%%%%%%%% BODY OF PAPER %%%%%%%%%%%%%%%%%%

\section{Introduction}

The past decade has seen a renaissance of the radio transient sky. While a number of transient and variable radio sources were known for many years from targeted searches of sources discovered at other observing frequencies, for example X-ray binaries (XRBs), active galactic nuclei (AGNs) and gamma-ray burst (GRB) afterglows, the typical radio transient sky was not well probed. The rapid development of new instrumentation has enabled us to conduct large scale surveys to systematically explore the radio transient sky over a range of timescales. For example, at high time resolution, typically $<$1 second, this led to the discovery of a new category of radio transient sources referred to as Fast Radio Bursts \citep[FRBs;][]{lorimer2007} that are pushing to the extremes of physics and enabling us to probe the contents of the Universe \citep[see][for a recent review of FRBs]{petroff2019}. 

At a lower time resolutions, typically $>$1 second, new wide field imaging software \citep[e.g., {\sc WSClean}; ][]{offringa2014}, telescopes with good instantaneous $uv$-coverage, and significant advances in computational power, have enabled large sky areas to be imaged on multiple timescales. This has led to a number of large surveys for transient and variable sources ranging from seconds \citep[e.g.,][]{kuiack2021} to tens of years \citep[e.g.,][]{radcliffe2019, bhandari2018} at a wide range of observing frequencies in the radio spectrum. Initially, these surveys were only capable of identifying bright sources in a large sky area \citep[e.g.,][]{bower2011} or faint sources in a much smaller sky area \citep[e.g.,][]{frail2012}. New facilities have now come online that can quickly survey large sky areas to high sensitivity. While transient radio astronomy focuses on a wide range of observing frequencies \citep[30 MHz -- 150 GHz; e.g.,][]{varghese2019,whitehorn2016}, in the following work we shall consider observations at $\sim$1.4 GHz to enable direct comparison to our observations. At mid-frequencies, the MeerKAT \citep[Meer Karoo Array Telescope;][]{camilo2018} and ASKAP \citep[Australian SKA Precursor;][]{hotan2021} radio telescopes have recently started observing with unprecedented sensitivity and rapid survey speeds, enabling deep probes of the radio transient sky on a wide range of timescales, leading to the discovery of transient and variable sources \citep[e.g.,][]{wang2021,driessen2020}.

Transient sources are typically considered to be those sources which appear and then disappear due to a cataclysmic event, while variable sources are those whose flux density varies between some minimum and maximum value. However, the observational division between transient and variable sources is more complex. For instance, a source that appears and disappears during an observation could be truly transient but, on the other hand, it could be a variable source with a minimum flux density that is below the detection threshold in the image. In this work, we define transient sources as those appearing and disappearing during the observations with the caveat that they may be previously unknown variable sources. We note that this bi-modal classification and the definition of transient versus variable sources are subject to debate within the literature. As we now have a growing population of these sources, we will likely move away from this bi-modal classification and towards a classification system based upon specific object types such as supernovae, AGN, etc.

While many unbiased transient surveys have been conducted at 1.4 GHz \citep[e.g.,][]{mooley2016,hodge2013,bell2011,bower2010}, only a small number of transient sources have been identified at 1.4 GHz \citep[e.g.,][see Table \ref{table:transientSDs} for more details]{levinson2002,thyagarajan2011,aoki2014}. \cite{levinson2002} identified 25 transient candidates on timescales of $\sim$1 year and they are believed to be orphan afterglows of radio afterglows from GRBs, radio supernovae and radio loud AGNs. \cite{thyagarajan2011} found 71 transient candidates on a wide range of timescales from 3 minutes up to 1 year using all the images obtained as part of the FIRST survey \cite{becker1995}. Many of the transient candidates have no known counterpart or identification, with the majority being consistent with galaxies and Quasi Stellar Objects. Finally, \cite{aoki2014} conducted a shallow survey for transients with flux densities greater than 3 Jy covering a very large area of sky and found one highly significant transient event on the timescale of 1 day. Thus, there are transient sources detected across a range of timescales at 1.4 GHz. Further observations are required to determine the nature of these sources and their populations.

Variable sources have proven to be more prolific at $\sim$1.4 GHz, with known sources being part of targeted monitoring campaigns \citep[e.g.,][]{bright2020} and the discovery of variable sources in wide field searches of survey datasets \citep[e.g.,][]{driessen2022b,murphy2021}. The variable sources discovered in wide field surveys vary on a wide range of timescales and include flaring stars \citep[e.g.,][]{driessen2022,driessen2020}, pulsars \citep[e.g.,][]{murphy2021}, tidal disruption events \citep[e.g.,][]{anderson2020} and AGNs \citep[e.g.,][]{murphy2021}. A summary of expected variable and transient sources on specific timescales is given in \cite{pietka2015}. Wide field surveys with multiple snapshots are ideal for identifying variable sources at 1.4 GHz, leading to the discovery of new variable sources and examples of rare extreme variability for known sources.

In this paper, we present a transient and variability search using observations obtained as part of the ThunderKAT XRB monitoring campaign \citep{fender2016}. Specifically, we used observations of the field of MAXI J1820+070 that were attained on a roughly weekly cadence \citep{bright2020}. In Section \ref{sec:obs_meerkat}, we outline the observational data used, the processing strategy and image quality control. In Section \ref{sec:hunting}, we outline the strategy and results for two optimised searches for transient and variable sources. Section \ref{sec:analysis} considers the progenitors of the identified transient and variable sources from this survey. 

Throughout this work, we adopt a cosmology with $H_{0} =71$ km s$^{-1}$ Mpc$^{-1}$, $\Omega_{m} = 0.27$ and $\Omega_{\lambda} = 0.73$.

\section{MeerKAT Observations}
\label{sec:obs_meerkat}

The observations used in this publication were obtained as part of the ThunderKAT weekly monitoring programme for XRBs. The target source, MAXI J1820+070, was observed weekly during its active phases \citep{bright2020}. Sixty four observations were obtained at a central frequency of 1.28 GHz, with a bandwidth of 0.86 GHz, and for a total integration time of 15 minutes per epoch. The observation dates are provided in Table \ref{table:observations}. These data were processed using the {\sc oxkat}\footnote{https://github.com/IanHeywood/oxkat} pipeline with standard parameters unless otherwise stated \citep{heywood2020}. \cite{heywood2022} describes the {\sc oxkat} pipeline in depth and it has three key stages. First, initial automatic flagging is conducted on the visibilities of the calibrator field using {\sc tricolour}\footnote{https://github.com/ratt-ru/tricolour} \citep{hugo2022}, followed by calibration of the data using both a primary and secondary calibrator. The primary calibrators used for these data, J1934-638 and PKS B1934-638, provided delay, bandpass, and frequency independent amplitude and phase solutions. The gain solutions are then transferred to the secondary calibrator, J1733-1304, which is used to calculate the complex, frequency dependent gain solutions. These solutions are then applied to the target data. The second phase of the {\sc oxkat} pipeline conducts automated flagging of the target using {\sc tricolour} and creates an initial image of the field and generates a mask. The third phase of {\sc oxkat} conducts a masked deconvolution of the target data and predicts the model visibilities that are used in a final self-calibration and imaging step. A single self-calibration cycle was conducted using {\sc CubiCal}  \citep[][]{kenyon2018} to obtain both phase and delay solutions. All imaging is conducted using {\sc WSClean} \citep{offringa2014}. The images per timestep are created at 4 frequency bands (0.96, 1.18, 1.39 and 1.61 GHz), together with a single combined multi-frequency image \citep[created using a fourth order spectral polynomial fit to the data, for more details see][]{offringa2017}. Each image is 10222$\times$10222 pixels, covering a field of view of 2.9$\times$2.9 square degrees. Observations that failed to be processed by this automated pipeline were excluded from the sample. Finally, a primary beam correction was applied to the images using the {\sc oxkat} tool {\sc pbcor\_katbeam.py}\footnote{https://github.com/ska-sa/katbeam} with image masking beyond the primary beam power point of 1 percent.

We note that there may be slight flux density variations between the images caused by offsets in the absolute image flux density scaling. To correct for this, we need to assume that, on average, the sources detected in the images do not have significant intrinsic variability. We chose the first successful image (Observation 6) at the highest observing frequency (1.61 GHz; as it has the highest resolution) and extracted the sources in that image using {\sc PySE} \citep{carbone2018}. We carefully choose sources to correct for the absolute flux scale offset using the following criteria.
\begin{itemize}
    \item We want to exclude any variations in the outer regions of the image, where the scaling may be affected by primary beam uncertainties. Thus, we extract all sources within 0.5 degrees of the image centre, minimising the primary beam uncertainties to 5 percent.
    \item Sources must be bright and significantly above the image noise. We use all sources detected above 20$\sigma$.
    \item Sources must be point sources, as extended sources can lead to flux density variations caused by source extraction methodology. The sourcefinder used in this analysis is {\sc PySE}, which is purposefully designed to handle point sources efficiently and is known to model extended sources less well. We chose the point sources by visual inspection of the sources in the image used to extract the sources.
\end{itemize}
Following these selection criteria, we obtain 13 point sources that can be used to correct for any systematic offsets in the absolute flux density scale of the image. We input the list of sources into the LOFAR Transients Pipeline \citep[{\sc TraP}; ][]{swinbank2015} using the monitoring list capability and ran the pipeline on all the images. The detection threshold was chosen such that no other sources were detected in the images. We take the average flux density of a source across the observations as the reference value to compare to the individual extractions of that source. For each image, we plot the extracted flux density of each source against its average flux density and fit a linear regression model through the origin with the {\sc Numpy} \citep{harris2020} least squares fitting algorithm. The gradient of the fitted linear model gives the flux density correction factor. The absolute flux density scale of the image is then corrected by dividing each pixel value by the flux density correction factor. This method is conducted for each observing frequency separately. The typical flux density correction factor is $\sim$4\%.

To assess image quality, we measure the rms noise variation in the inner eighth of the images. We fit a Gaussian distribution to the rms noise values to obtain the average rms value and the 1$\sigma$ uncertainty. In Table \ref{table:image_noise}, we give the average rms noise for the sample of images at each observing frequency. Images with rms noise values that are $>3\sigma$ deviant from the average rms value are rejected. Additionally, images where the restoring beam ellipticity is $>2$ are rejected as this can cause issues in source association within the {\sc TraP}. Finally, images are checked by visual inspection. The images at 1.18 GHz and 1.61 GHz from Observation 29, on 2019 May 18, are rejected due to having a very poor resolution. The images rejected are given in Table \ref{table:observations}. All other images are deemed as being of suitable quality for transient and variability hunts.

\begin{table}
\centering
\begin{tabular}{|c c c|} 
\hline
Frequency band & average rms & Number of images\\
(GHz) & ($\mu$Jy beam$^{-1}$) & \\
\hline
0.96 &  83.8$^{+7.5}_{-6.8}$   & 53 \\
1.18 &  72.0$^{+19.0}_{-15.0}$ & 62 \\
1.39 &  42.7$^{+2.8}_{-2.6}$   & 47 \\
1.61 &  75.9$^{+32.6}_{-22.8}$ & 62 \\
\hline
\end{tabular}
\caption{The typical rms noise in the inner eighth of the images at each observing band. The average rms values are obtained by fitting a Gaussian to the rms distribution in logarithmic space and the uncertainties are the 1$\sigma$ values. The number of images per observing frequency are those images that have passed the quality control and are processed using the {\sc TraP}.}
\label{table:image_noise}
\end{table}

\section{Transient and Variable hunt}
\label{sec:hunting}

We input the images attained in Section \ref{sec:obs_meerkat} directly into {\sc TraP} using default settings except for the parameters given in Table \ref{table:TraP}. The {\sc TraP} takes a time series of images and conducts basic quality control on these images prior to further processing. After the quality control step, {\sc TraP} searches for all sources in the images using the {\sc PySE} source finder. Sources are then associated across time and frequency to produce multi-wavelength light curves. If a source is not detected, a constrained fit of the flux density is obtained at the location of the source. Finally, variability metrics are calculated for the output light curves \citep{swinbank2015}.

The detection\_threshold parameter determines the signal to noise ratio used by the source finder to detect sources within the images; in this analysis we use different values for the transient and variability analysis in order to optimise those searches (see the following subsections). The extraction\_radius parameter is the radius (in pixels) out to which sources are found by the source finder. The force\_beam parameter, when set to true, assumes all sources are point sources as expected for transient sources on the timescales and resolutions probed by this analysis. The new\_source\_sigma\_margin parameter can be used to raise the detection threshold so that new sources need to be more significantly detected before they are labelled as being transient. In this analysis, we choose to turn off the new\_source\_sigma\_margin and instead filter transient candidates via other methods (see the following subsections). The elliptical\_x parameter is used to filter images of lower quality; in our analysis any images where the restoring beam is significantly elliptical (i.e., $B_{\rm major}/B_{\rm minor} > 2$) are rejected from the analysis. Finally, the beamwidths\_limit parameter is used to control how far, in units of the restoring beam, from a newly detected source the source association algorithm can search for associations.

Following the quality control steps outlined in the previous sections, 3--27 percent of images were rejected. All of the rejected images were due to having high image noise and none were rejected due to the ellipticity of the restoring beam. In Table \ref{table:image_noise}, we provide the number of images per observing frequency that remain in the final sample.

\begin{table}
\centering
\begin{tabular}{|c c|} 
\hline
Parameter & Value \\
\hline
detection\_threshold     & 4 (transient hunt; \ref{sec:transient_hunt}) \\ %5.56
                         & 8 (variability hunt; \ref{sec:variable_hunt}) \\
extraction\_radius       & 2945 pixels (0.9 degrees)\\
force\_beam              & True \\
new\_source\_sigma\_margin & 0 \\
elliptical\_x & 2 \\
beamwidths\_limit     & 1 (transient hunt; \ref{sec:transient_hunt}) \\
                         & 3 (variability hunt; \ref{sec:variable_hunt}) \\
\hline
\end{tabular}
\caption{The parameters used within the {\sc TraP}. The optimal detection threshold is calculated in Section \ref{sec:transient_hunt}.}
\label{table:TraP}
\end{table}

\begin{figure}
\centering
\includegraphics[width=0.48\textwidth]{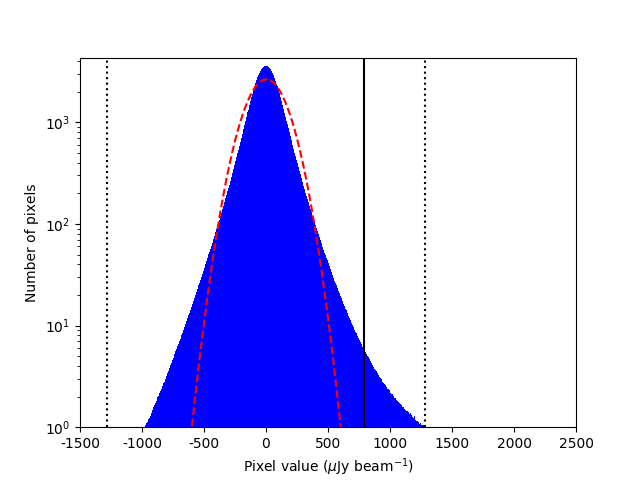}
\caption{Histogram of all the pixel values searched for transient sources for the images obtained at 0.96 GHz. The black dotted lines show the median clipping threshold conducted on the pixels and it has a negligble impact on the results. The red dashed line shows the best fit Gaussian distribution for this observing frequency, and the black solid line shows the optimal detection threshold of $5.22\sigma$ for this dataset.}
\label{fig:pixels}
\end{figure}

\begin{table*}
\centering
\begin{tabular}{|c c c c c|} 
\hline
Frequency & pix beam$^{-1}$ & Mean & Sigma & Detection threshold \\
(GHz) &  & (mJy beam$^{-1}$) & (mJy beam$^{-1}$) & ($\sigma$) \\
\hline
0.96 & 121 & $2.54\times 10^{-3}$ & 0.15 & 5.22 \\
1.18 & 59  & $1.58\times 10^{-3}$ & 0.19 & 5.38 \\
1.39 & 40  & $8.39\times 10^{-4}$ & 0.15 & 5.40 \\
1.61 & 21  & $5.63\times 10^{-4}$ & 0.33 & 5.56 \\
\hline
\end{tabular}
\caption{The pixel properties for all the images at each observing band. The mean and sigma parameters are from a Gaussian distribution fitted to the pixel values of all the pixels at each observing band following sigma clipping to remove sources. The optimal detection threshold is the $\sigma$ value chosen to ensure that {\sc TraP} detects no more than one false positive transient source per frequency band.}
\label{table:image_pixels}
\end{table*}

\subsection{Transient hunt}
\label{sec:transient_hunt}

\begin{figure*}
\centering
\includegraphics[width=0.95\textwidth]{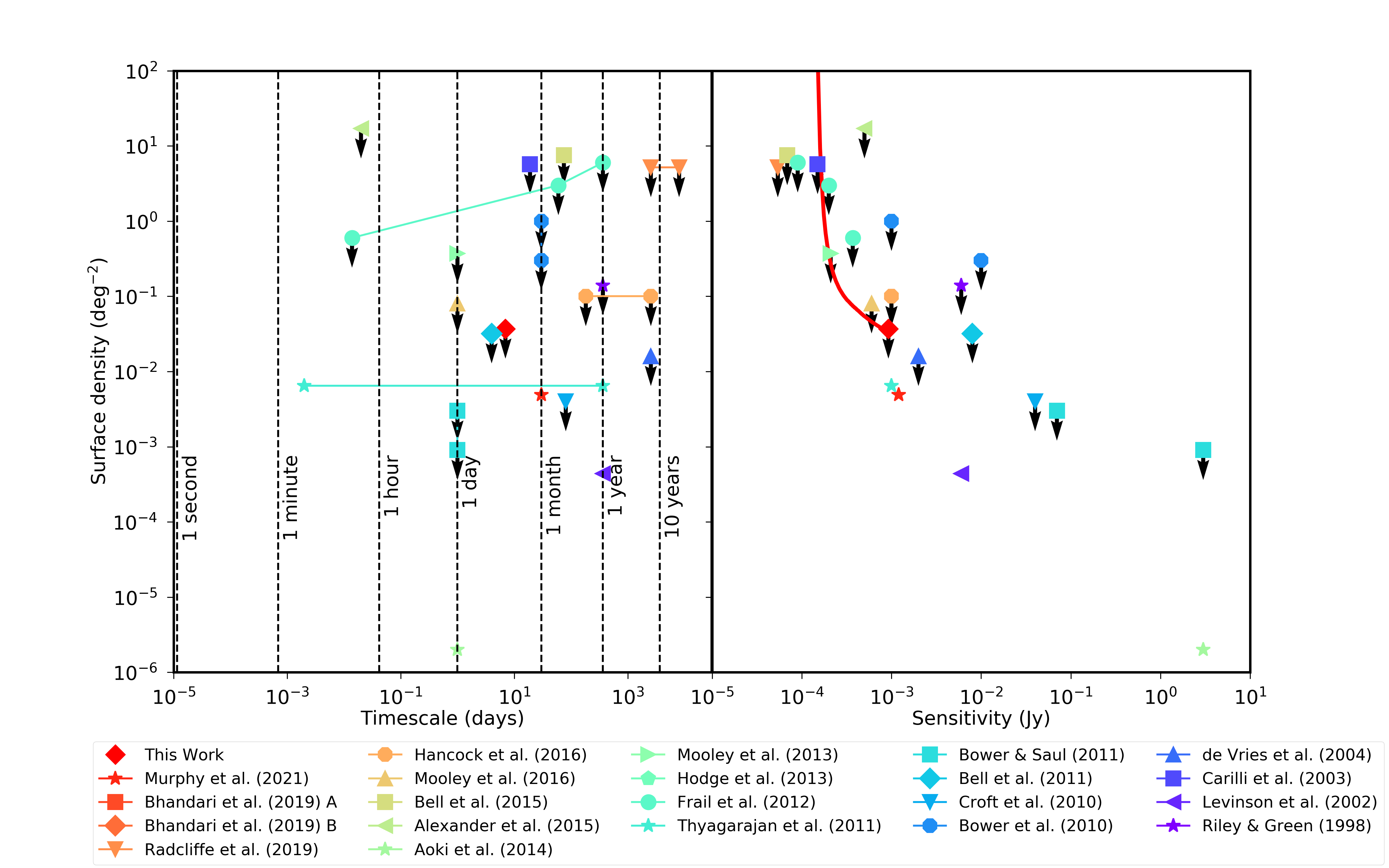}
\caption{The transient surface density limits for a range of transient surveys conducted at 1.4 GHz. The right hand panel shows the transient surface density as a function of the faintest transient detectable (defined as the sensitivity). The left hand panel shows the transient surface density as a function of the timescale between each consecutive snapshot image used. Data points with arrows are upper limits on the transient surface density and data points without arrows represent surveys that detected transient sources. The red thick line and red diamond data points represent the results attained in this work at the observing frequency of 1.39 GHz. The data plotted in this Figure are presented in Table \ref{table:transientSDs}.}
\label{fig:surfaceDensityPlot}
\end{figure*}

%Assuming binomial statistics as there are two outcomes: transient and not transient
%In binomial statistics, we can calculate the maximum likelihood parameter by $p = \frac{\rm no.~succcess}{\rm no.~trials}$.
%In this case, we want 1 ``success''. The number of trials is given by $\frac{\rm total~pixels}{\rm pixels~in~beam}$. Thus $p = \frac{1}{N} = 1 / \frac{\rm total~pixels}{\rm pixels~in~beam}$
%But actually, we don't want this outcome, we want the probability of failure. i.e., $q=1-p = 1 - \frac{1}{N}$
%Then we fit a normal distribution to the data and want to find the pixel value where the probability contained on the left hand side is equal to the probability calculated via binomial statistics. This is calculated using norm in scipy stats. The number of sigma is then given by the pixel value divided by the standard deviation of the normal distribution.

Transient sources are defined as those sources that are newly detected during the timescale of the observations. To determine a reliable detection threshold for new sources, we require to detect less than 1 false positive source in all the images we process.  Due to correlated noise in radio interferometers, features in the images take the shape of the restoring beam. Thus, we need to calculate the false positive rate by using the number of independent beam elements in the image rather than the number of pixels. The probability of this occurring is given by $P(X\le 1) = 1 - \frac{1}{N}$ (assuming a binomial distribution and taking the maximum likelihood of there being one false detection), where the number of trials, $N$ is given by the total number of pixels divided by the number of pixels in one restoring beam. We extract the pixel values for all the pixels contained within the extraction\_radius used by {\sc TraP} (given in Table \ref{table:TraP}) and plot them in a histogram. In Figure \ref{fig:pixels}, we show an example histogram for the images obtained at a frequency of 0.96 GHz. The dataset contains $2.5\times 10^{7}$ pixels per image, corresponding to $\sim 1.6 \times 10^{9}$ pixels per frequency band.

To calculate the optimal detection threshold for each observing frequency, we assume that the noise pixels follow a Gaussian distribution. We note that the distribution does not follow a Gaussian distribution, this is due to an excess of positive and negative sidelobes around bright sources in the outer regions of the image. Direction dependent calibration can significantly reduce these sidelobes, however we chose not to do this as it could not be fully automated, it takes significantly longer to process and the sidelobes can be easily distinguished from transient candidates (see Section \ref{sec:transient_hunt}). Even though these data do not follow a Gaussian distribution, this assumption is good enough to set a threshold for the initial search for transients. The pixel values also contain real sources detected in the images, thus we conduct clipping of the pixel data to remove those sources. We calculate the median value of the pixels and then remove all pixels that are outside of two standard deviations from the median. After clipping, the remaining noise pixels are fitted with a Gaussian distribution and we show the parameters in Table \ref{table:image_pixels}. Using the Gaussian distributions, we calculate the sigma detection threshold required to provide $P(X\le 1)$. To be conservative in our transient identification, we use the maximum detection threshold from the four observing frequencies. We find that the detection threshold providing less than 1 false positive for all observing frequencies should be $5.56\sigma$.

To search for new sources, we run {\sc TraP} with a 4$\sigma$ detection threshold on all of the images at once in order to obtain a list of all new sources detected throughout the dataset. For this low detection threshold, the source association becomes increasingly complex and can fail. Therefore, we significantly simplify the source association to sources within 1 restoring beam width of each other in the image being processed. This simplification may lead to missed source associations for sources within the image but this does not hinder the transient search due to the filtering strategy outlined below. The number of new sources (those not detected in the first time step processed) detected is 42090, across the four observing bands. We note that this number does not exclude multiple detections of the same source across the frequency bands. These sources are filtered using the following criteria.

\begin{enumerate}
    \item Extract only sources from the new source list that are detected with a signal to noise ratio above the calculated detection threshold of 5.56$\sigma$. This reduces the number of new sources by 99\% to 334 remaining.
    \item Reject any sources that can be associated with existing sources in the database. These are likely caused by source association issues within the pipeline. This results in a reduction of new sources by 31\%, bringing the candidate list to 231 sources.
    \item Reject any sources close to the source extraction radius. The {\sc TraP} measures the noise within the extraction radius by subdividing the area into smaller blocks and extrapolating the noise between these blocks \citep[see][for further details]{swinbank2015}. This can lead to an underestimate of the noise around the source extraction radius and thus a number of false positive transient detections \citep[e.g.,][]{rowlinson2016}. We reject any new sources identified within 3 arcminutes of the source extraction radius. This removes 9\% of candidates remaining from the second filtering step, leading to 210 sources remaining on the candidate list.
    \item Reject any candidates only detected in one frequency band. Often in radio images there are noise features surrounding bright sources, known as sidelobes, which can be mistaken for new sources. Additionally, correlated noise features can also lead to false positive source detections. These noise features are an artefact of the images being made via an incomplete Fourier transformation on sparsely sampled data. By conducting self-calibration, as with the data presented in this paper, it is possible to significantly reduce these artefacts but some remain. The locations of these noise features are highly dependent on the observing frequency; meaning that one of these false positive results will not have a corresponding source at the same time and location at a different observing frequency. Thus, we require that all new transient candidates are detected in at least 2 frequency bands in the same timestep. At least one detection must exceed the detection threshold of 5.56$\sigma$ and at least two detections must exceed the 4$\sigma$ detection threshold used in TraP. This reduces the number of candidates remaining from the previous step by 98\%, leading to a total of 4 candidates.
    \item Confirm that candidates are not underlying faint sources near the detection threshold used in this analysis. We create deep images for a given observing frequency by summing all the pixels in each image obtained and dividing the resultant pixel values by the number of images used (note, this is not as good as re-imaging all of the calibrated visibilities together but is sufficient for this analysis). To confirm the flux density scale in these averaged images is reasonable, we plot the observed flux density in the mean image against the average flux density measured by the TraP for all sources with a detection significance above 8$\sigma$, and there is no significant systematic offset between the fluxes. We then reject any candidates detected in the deep image for the detection frequency where the candidate maximum flux density is more than 5 $\sigma$ deviant from the flux density measured in the deep image. Sources not associated with a source in the deep image are also retained as transient candidates. This removes none of candidates from the previous filtering step, leading to 4 sources remaining on the candidate list. 
    \item Conduct visual inspection of all remaining candidates using both the individual images and the deep image of the field. This leads to the rejection of one candidate as it is comparable to noise features in the nearby region.
    \item Run {\sc TraP} with the remaining three candidates in a monitoring list to obtain a full light curve and the variability parameters. Their properties are given in Table \ref{table:variable_candidates}.
\end{enumerate}

%\begin{table*}
%\centering
%\begin{tabular}{|c c c c c c c c c c l|} 
%\hline
%ID & RA & Dec & RA error & Dec error & Frequency & Max Flux & Max Flux error & $V$ & $\eta$ &  Comment \\
% & (degrees) & (degrees) & (arcsec) & (arcsec) & (GHz) & (mJy) & (mJy) & & & \\
%\hline
%1  & 275.5770 & 7.0885 & 1.091 & 1.083 & 0.96 & 1.9 & 0.2 & 0.535 & 1.866 &  PSR J1822+0705 \\ % 17644
% & & & & & 1.18 & 1.6 & 0.2 & 0.467 & 1.013 & \\ % monitor 59613
% & & & & & 1.39 & 0.5 & 0.1 & 0.428 & 0.602 & \\
% & & & & & 1.61 & 0.4 & 0.1 & 0.952 & 0.538 & \\
%\hline
%2   & 275.0663 & 7.2485 & 2.544 & 2.524 & 0.96 & 0.4 & 0.1 & 0.618 & 0.771 & MKT J182015.5+071455 \\ % 20764
% & & & & & 1.18 & 0.3 & 0.1 & 0.702 & 0.524 & \\ % monitor 56914
% & & & & & 1.39 & 0.2 & 0.1 & 0.376 & 0.391 & \\
% & & & & & 1.61 & 0.2 & 0.1 & 0.570 & 0.315 & \\
%\hline
%3   & 275.0060 & 7.4958 & 1.891 & 1.875 & 0.96 & 0.5 & 0.2 & 0.443 & 0.558 & MKT J182001.4+072945 \\ % 25568. PSO J275.0057+07.4958
% & & & & & 1.18 & 0.4 & 0.1 & 0.507 & 0.412 & \\ % monitor 59615
% & & & & & 1.39 & 0.3 & 0.1 & 0.297 & 0.345 & \\
% & & & & & 1.61 & 0.5 & 0.2 & 0.779 & 0.625 & \\
%\hline
%\end{tabular}
%\caption{The transient candidates identified in the field of MAXI J1820+070.}
%\label{table:transient_candidates}
%\end{table*}

\begin{figure}
\centering
\includegraphics[width=0.48\textwidth]{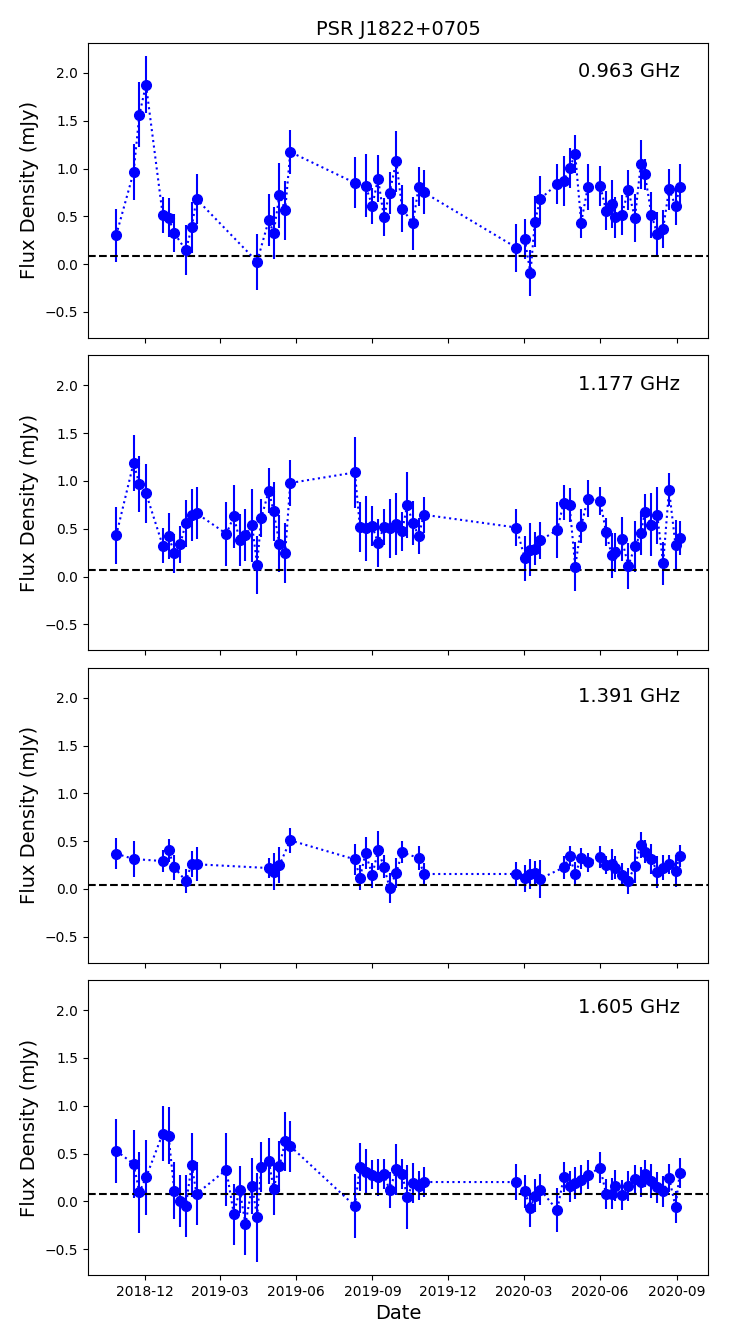}
\caption{The observed flux density light curve of PSR J1822+0705, identified as a variable source in Section \ref{sec:transient_hunt}. The dashed horizontal line represents the typical rms noise in the images as given in Table \ref{table:image_noise}.}
\label{fig:trans_lightcurves1}
\end{figure}

\begin{figure}
\centering
\includegraphics[width=0.48\textwidth]{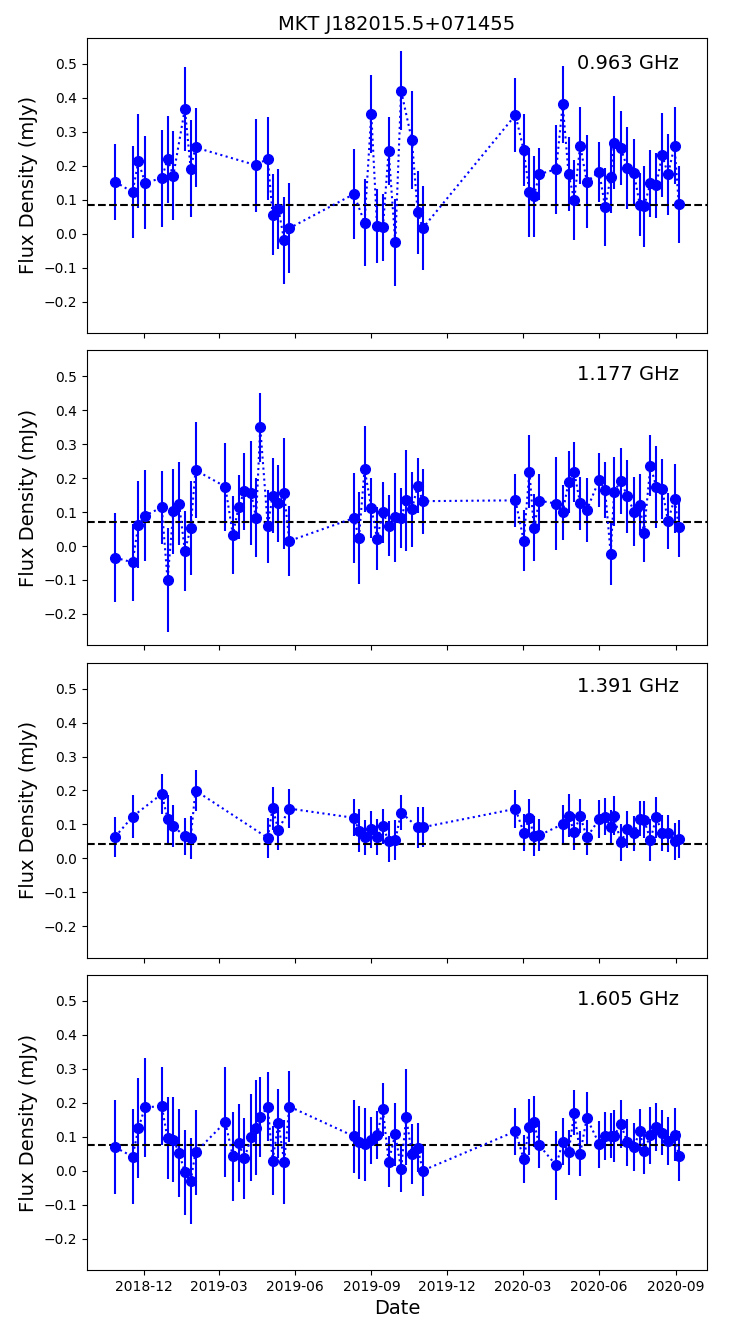}
\caption{The observed flux density light curve of MKT J182015.5+071455, identified as a variable source in Section \ref{sec:transient_hunt}. The dashed horizontal line represents the typical rms noise in the images as given in Table \ref{table:image_noise}.}
\label{fig:trans_lightcurves2}
\end{figure}

\begin{figure}
\centering
\includegraphics[width=0.48\textwidth]{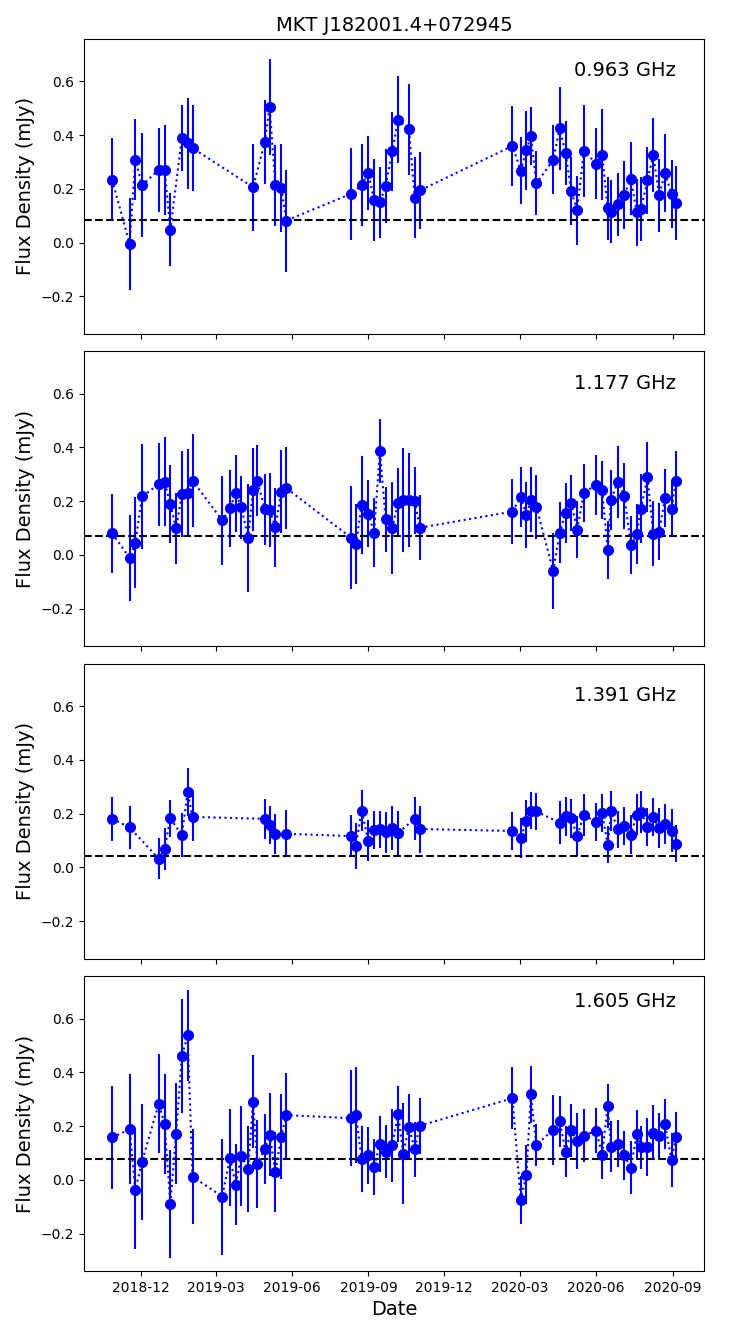}
\caption{The observed flux density light curve of MKT J182001.4+072945, identified as a variable source in Section \ref{sec:transient_hunt}. The dashed horizontal line represents the typical rms noise in the images as given in Table \ref{table:image_noise}.}
\label{fig:trans_lightcurves3}
\end{figure}

In summary, three candidate sources pass these criteria, although all of them are associated with sources in the deep images of the field and they are defined to be variable sources. Their light curves are shown in Figures \ref{fig:trans_lightcurves1} -- \ref{fig:trans_lightcurves3}. The method outlined above shows an efficient method to automatically identify reliable transient candidates in a fully automated manner using the standard outputs of the {\sc TraP}.

%\begin{figure*}
%\centering
%\includegraphics[height=0.24\textheight]{59613_lightcurve.png}
%\includegraphics[height=0.24\textheight]{59614_lightcurve.png}
%\includegraphics[height=0.24\textheight]{59615_lightcurve.png}
%\caption{The observed flux density light curves of the 3 sources identified as variable sources in Section \ref{sec:transient_hunt}.}
%\label{fig:trans_lightcurves}
%\end{figure*}

\subsubsection{Transient surface density}

The sources detected in this analysis are all associated with faint sources in the deep images and are observed in multiple epochs. These are identified as variable sources rather than the transient sources typically considered in standard transient searches. Therefore, we do not include these sources in comparisons to other transient surveys. In order to compare this transient search to other published searches, we need to determine the transient surface density at a single observing frequency. We use the data presented in Table \ref{table:transientSDs} and obtained from \cite{mooley2016}\footnote{http://www.tauceti.caltech.edu/kunal/radio-transient-surveys/index.html}. We choose to use the images observed at 1.39 GHz as these images were the best quality and we are comparing this result to other surveys conducted at 1.4 GHz. The transient surface density is defined as being the number of transients detected per square degree surveyed and is typically compared to the sensitivity of the observations (the faintest transient detectable in the survey). 

The images typically have structure in their noise properties; for instance the quality decreases with increasing distance from the image centre due to the response of the primary beam. Thus, we sample a small area to high sensitivity and a significantly larger area to a lower sensitivity limit. 

Following a similar strategy to that outlined by \cite{kuiack2021}, we divide each image into 100 annuli, determine the sky area covered by each annulus, and then calculate the rms noise in each annulus (following a simple sigma clipping, with a 3$\sigma$ threshold, to remove bright sources from the data). Following this step, we determine the maximum and minimum rms values and create a range of values between this limit with 100 bins. The sky area surveyed by each unique annulus within a given rms bin is then summed up. Finally, for each rms bin we sum up the total sky area surveyed for all rms bins less than or equal to this bin. Hence, the final bin, at the highest rms value, has the total area surveyed in all the images. As we search out to a radius of 0.9 degrees, the area surveyed in one image is equal to $\Omega = 0.9^2 \pi$ deg$^{2}$. Thus, the total area surveyed is $0.9^2 \pi \times (N-1)$ deg$^{-2}$, where $N$ is the number of images surveyed. The `$-1$' in this expression is because one of the images is needed as a comparison image and hence not included in the total area. For large datasets, $(N-1)$ tends towards $N$, therefore the summation method used above will tend towards the total area surveyed. 

To convert this to a transient surface density $\rho$, the number of transients $T$ found is divided by the total area surveyed, i.e., $\rho = \frac{T}{\Omega (N-1)} = \frac{T}{A}$ where $A$ is the area calculated via the binning strategy outlined above. Following previous studies, we calculate the 95\% confidence limit that no transient sources were detected assuming Poissonian statistics \citep[e.g.,][]{rowlinson2016,bell2014}, and this equation approximates to $\rho \simeq \frac{3}{A}$. The transient surface density is then calculated for each rms bin. To determine the sensitivity, or the faintest transient detectable, we multiply the rms value by the detection threshold used, i.e., $\sigma \times {\mathrm{rms}}$ where $\sigma = 5.56$ for this work. 

Finally, we plot the transient surface density as a function of the sensitivity in the right hand panel of Figure \ref{fig:surfaceDensityPlot} using the thick red line. The diamond at the end of this line represents the faintest transient detectable in the total area surveyed (i.e., the lowest transient surface density probed). In the left hand panel of Figure \ref{fig:surfaceDensityPlot}, we plot the lowest value of the transient surface density probed versus the median minimum separation timescale of the images in days (7 days for these data). We find a transient surface density of $<3.7 \times 10^{-2}$ deg$^{-2}$ for transients brighter than 0.1 mJy on timescales of 1 week.

\subsection{Variable hunt}
\label{sec:variable_hunt}

\begin{table}
\centering
\begin{tabular}{|c c c|} 
\hline
Frequency band & $\eta$ Threshold & $V$ Threshold \\
(GHz) &  &  \\
\hline
0.96 & 8.22  & 0.203 \\
1.18 & 16.5  & 0.196 \\
1.39 & 13.5  & 0.204 \\
1.61 & 50.2  & 0.179 \\
\hline
\end{tabular}
\caption{The $1.5\sigma$ thresholds for the variability parameters, $\eta$ and $V$, used for each observing frequencies.}
\label{table:variability_limits}
\end{table}

\begin{figure*}
\centering
\includegraphics[width=0.94\textwidth]{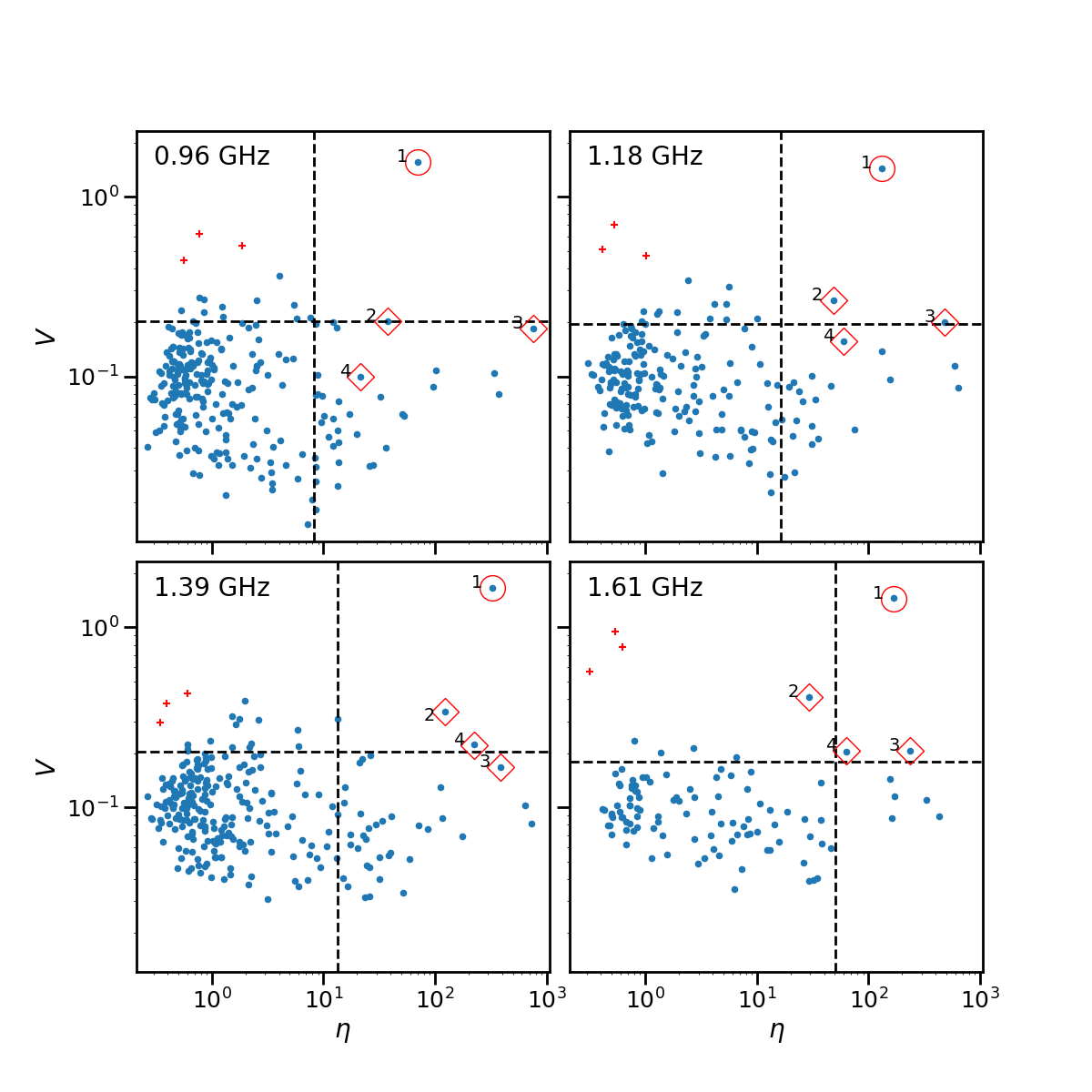}
\caption{The distributions of the two variability parameters, $\eta$ and $V$, for all unique sources detected in the MAXI J1820+070 field for each of the 4 observing frequencies. The dashed black lines show the $1.5\sigma$ thresholds used for filtering these parameters. All sources in the top-right regions of each plot are classed as variable candidates. Unsurprisingly, MAXI J1820+070 is clearly identified as a variable source and is highlighted with the red circle. All candidate variable source candidates are listed in Table \ref{table:variable_candidates}. The red plus markers show the candidate variable sources identified during the search for transient sources (see Section \ref{sec:transient_hunt}) and their properties are listed at the top of Table \ref{table:variable_candidates}.}
\label{fig:VandEta}
\end{figure*}

\begin{table*}
\centering
\begin{tabular}{|c c c c c c c c c l c |} 
\hline
\hline
ID & RA & Dec & Position error & Frequency & Avg Flux Density & Max Flux Density & $V$ & $\eta$ &  Source ID & Variable? \\
 & (degrees) & (degrees) & (arcsec) & (GHz) & (mJy) & (mJy) & & & & \\
\hline
\hline
\multicolumn{11}{l}{Identified in Section \ref{sec:transient_hunt}} \\
\hline
1  & 275.5770 & 7.0885 & 1.1 & 0.96 & & 1.9$\pm$0.2 & 0.535 & 1.866 &  PSR J1822+0705  &\\ % 17644
 & & & & 1.18 & & 1.6$\pm$0.2 & 0.467 & 1.013 &  &\\ % monitor 59613
 & & & & 1.39 & & 0.5$\pm$0.1 & 0.428 & 0.602 &  &\\
 & & & & 1.61 & & 0.4$\pm$0.1 & 0.952 & 0.538 &  &\\
\hline
2   & 275.0663 & 7.2485 & 2.5 & 0.96 & & 0.4$\pm$0.1 & 0.618 & 0.771 & MKT J182015.5+071455  &\\ % 20764
 & & & & 1.18 & & 0.3$\pm$0.1 & 0.702 & 0.524 &  &\\ % monitor 59614
 & & & & 1.39 & & 0.2$\pm$0.1 & 0.376 & 0.391 &  &\\
 & & & & 1.61 & & 0.2$\pm$0.1 & 0.570 & 0.315 &  &\\
\hline
3   & 275.0060 & 7.4958 & 1.9 & 0.96 & & 0.5$\pm$0.2 & 0.443 & 0.558 & MKT J182001.4+072945  &\\ % 25568. PSO J275.0057+07.4958
 & & & & 1.18 & & 0.4$\pm$0.1 & 0.507 & 0.412 &  &\\ % monitor 59615
 & & & & 1.39 & & 0.3$\pm$0.1 & 0.297 & 0.345 &  &\\
 & & & & 1.61 & & 0.5$\pm$0.2 & 0.779 & 0.625 &  &\\
\hline
\hline
\multicolumn{11}{l}{Identified in Section \ref{sec:variable_hunt}} \\
\hline
1   & 275.0913 & 7.1854 & 2.7 & 0.96 & 0.6 & 3.2$\pm$0.1 & 1.56 & 69.4 & MAXI J1820+070 & \checkmark \\ % 3801
    &  &  &  & 1.18 & 0.8 & 4.0$\pm$0.1 & 1.43 & 131 &  & \checkmark \\ % 4246
    &  &  &  & 1.39 & 0.6 & 3.6$\pm$0.1 & 1.66 & 325 &  & \checkmark \\ % 4693
    &  &  &  & 1.61 & 0.8 & 4.2$\pm$0.1 & 1.45 & 168 &  & \checkmark \\ % 5076
\hline
2   & 274.7075 & 6.4790 & 1.4 & 0.96 & 7.4 & 10.2$\pm$0.2 & 0.203 & 37.5 & NVSS J181849+062843 & \checkmark \\ % 3723 
    &  &  &  & 1.18 & 9.9 & 15.1$\pm$0.4 & 0.265 & 48.8 &  & \checkmark \\ % 4189
    &  &  &  & 1.39 & 10.4 & 24.0$\pm$0.4 & 0.340 & 123 &  & \checkmark \\ % 4601
    &  &  &  & 1.61 & 13.4 & 32.7$\pm$1.3 & 0.410 & 29.3 &  &  \\ % 5047
\hline
3   & 274.4692 & 6.7773 & 1.3 & 0.96 & 102.7 & 132.9$\pm$1.0 & 0.185 & 756 & NVSS J181752+064638 &  \\ % 3908 
    &  &  &  & 1.18 & 115.0 & 150.0$\pm$1.5 & 0.201 & 484 &  & \checkmark \\ % 4161
    &  &  &  & 1.39 & 112.6 & 143.0$\pm$1.3 & 0.167 & 384 &  & \\ % 4565
    &  &  &  & 1.61 & 111.0 & 161.8$\pm$3.0 & 0.206 & 235 &  & \checkmark \\ % 5035
\hline
4   & 275.1254 & 6.5720 & 1.5 & 0.96 & 7.7 & 9.2$\pm$0.2 & 0.0992 & 21.6 & NVSS J182029+063419 & \\ % 3814 
    &  &  &  & 1.18 & 9.0 & 12.0$\pm$0.2 & 0.157 & 60.1 &  & \\ % 4258
    &  &  &  & 1.39 & 9.4 & 13.0$\pm$0.2 & 0.222 & 225 &  & \checkmark \\ % 4708
    &  &  &  & 1.61 & 10.1 & 13.2$\pm$0.2 & 0.205 & 63.2 &  & \checkmark \\ % 5084
\hline
\hline
\end{tabular}
\caption{The variable sources identified in the field of MAXI J1820+070.}
\label{table:variable_candidates}
\end{table*}

%\begin{figure*}
%\centering
%\includegraphics[height=0.24\textheight]{3723_lightcurve.png}
%\includegraphics[height=0.24\textheight]{3908_lightcurve.png}
%\includegraphics[height=0.24\textheight]{3814_lightcurve.png}
%\caption{The observed flux density light curves of the 3 sources identified as variable in Section \ref{sec:variable_hunt}}
%\label{fig:lightcurves}
%\end{figure*}

\begin{figure}
\centering
\includegraphics[width=0.48\textwidth]{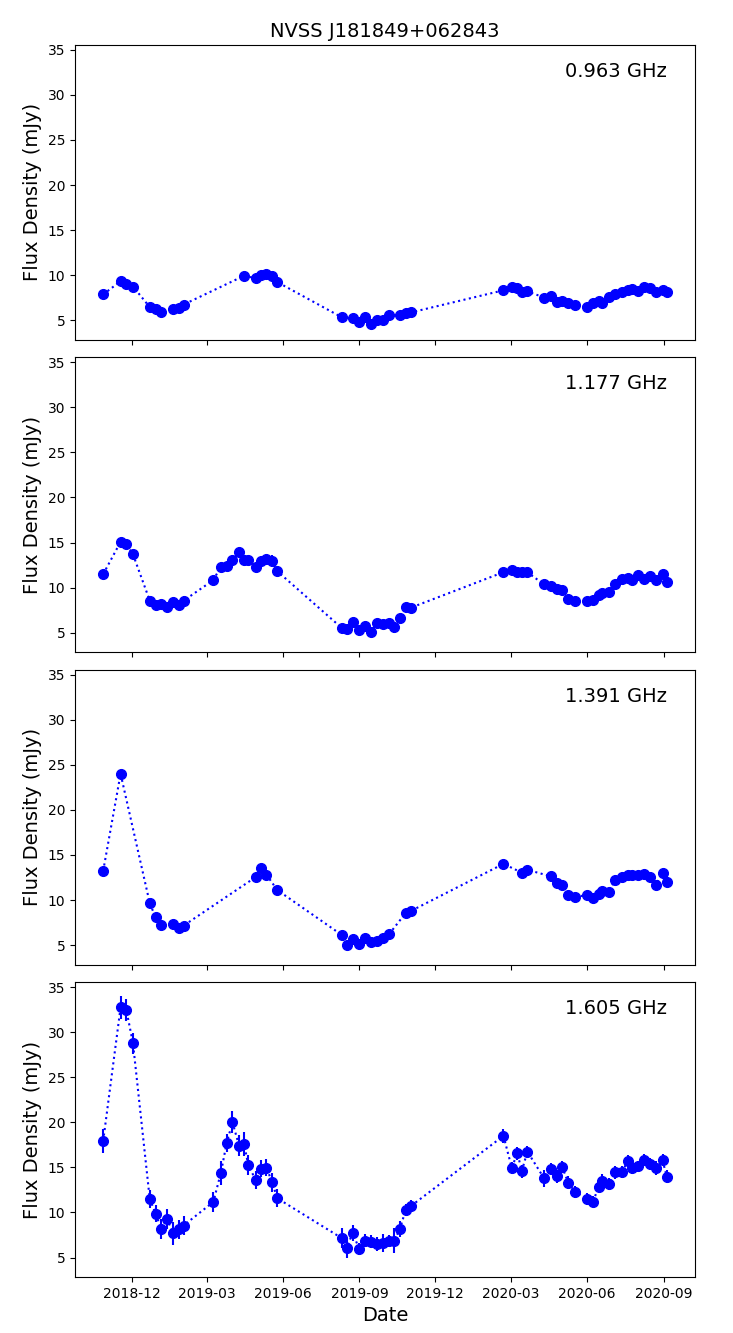}
\caption{The observed flux density light curve of NVSS J181849+062843, identified as a variable in Section \ref{sec:variable_hunt}.}
\label{fig:lightcurves1}
\end{figure}

\begin{figure}
\centering
\includegraphics[width=0.48\textwidth]{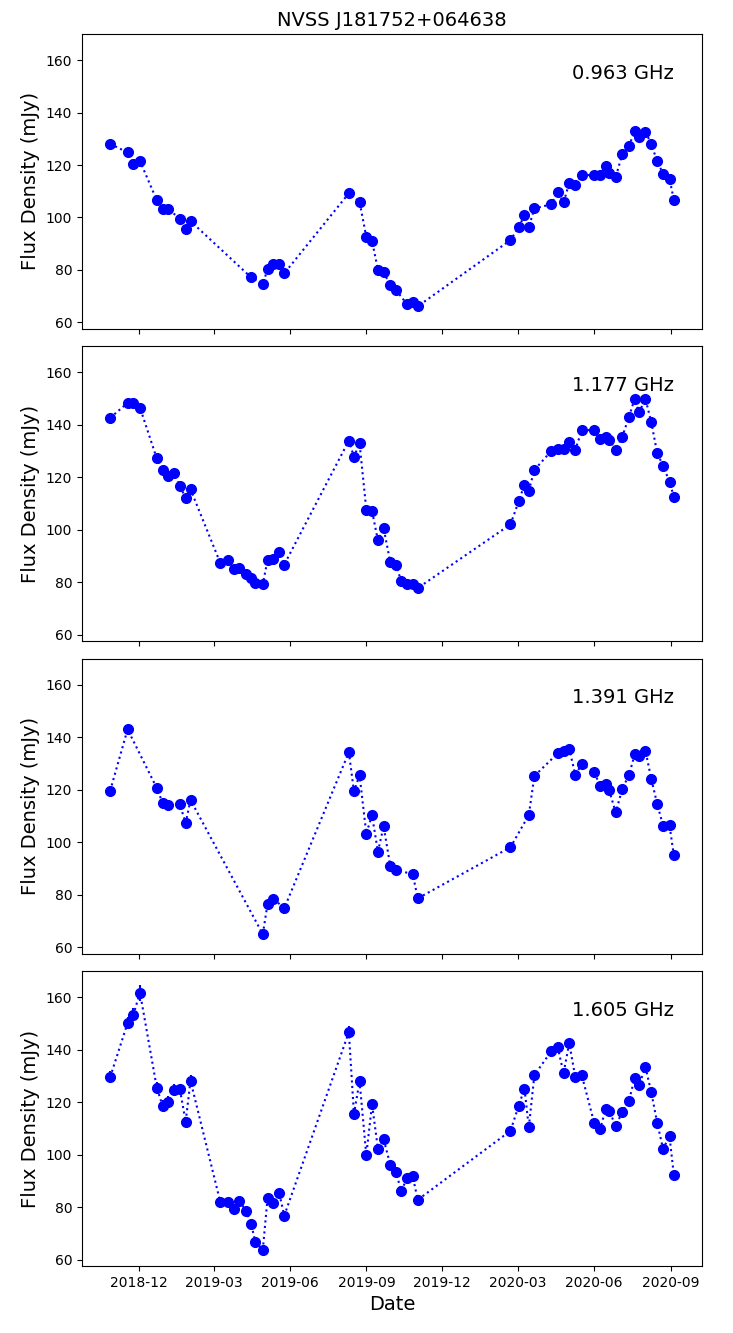}
\caption{The observed flux density light curve of NVSS J181752+064638, identified as a variable in Section \ref{sec:variable_hunt}.}
\label{fig:lightcurves2}
\end{figure}

\begin{figure}
\centering
\includegraphics[width=0.48\textwidth]{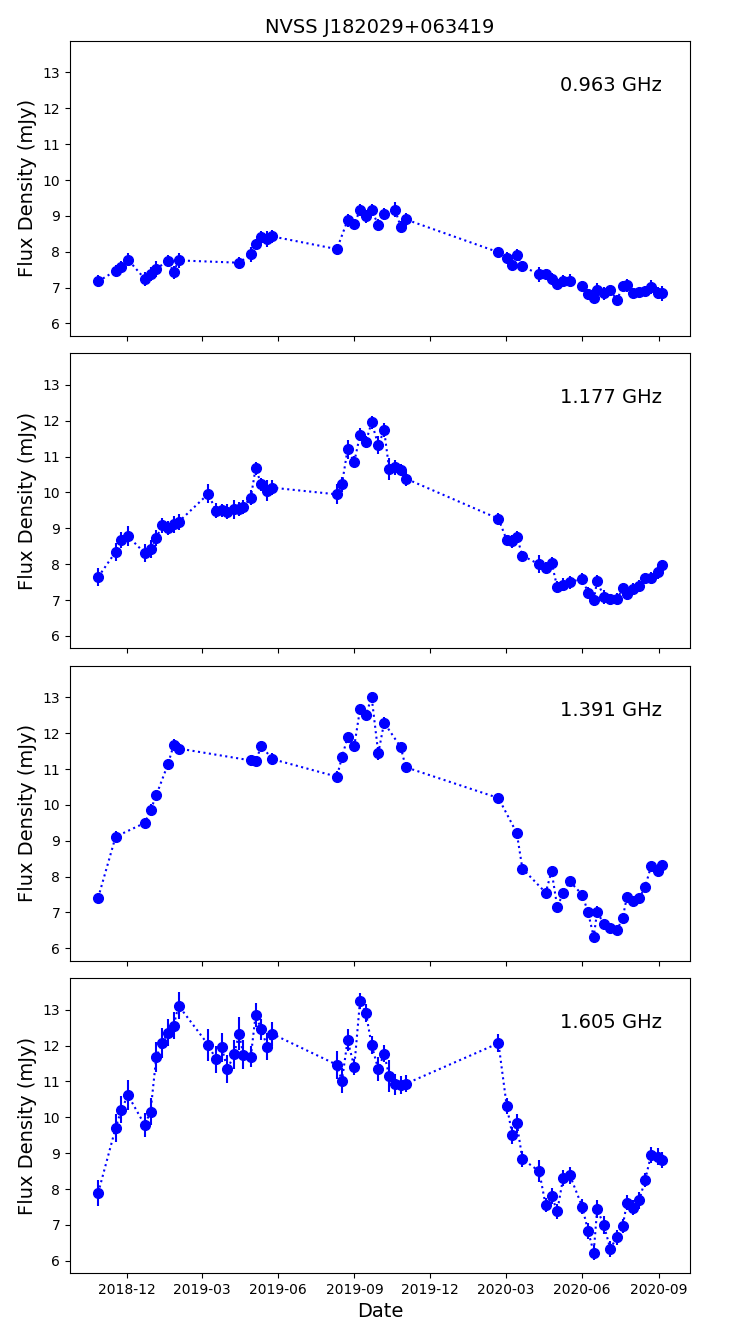}
\caption{The observed flux density light curve of NVSS J182029+063419, identified as a variable in Section \ref{sec:variable_hunt}.}
\label{fig:lightcurves3}
\end{figure}

In this analysis, variable sources are defined as those sources being detected in the first observation of the field and showing significant variation in their flux densities throughout the observing campaign. As variability is difficult to quantify for faint sources, we use a higher detection threshold than in the previous section, i.e., 8$\sigma$. Additionally, as we are concerned with variability of sources, it is important to not miss source associations. Using a higher detection threshold reduces the complexity of the source association procedure and we can use the default value of 3 for the {\sc TraP} beam widths limit to prevent missed source associations. Variable sources are typically identified using the two key variability parameters calculated by the {\sc TraP}: the reduced weighted $\chi^{2}$ of a fit assuming a constant brightness, $\eta$, and the coefficient of variation, $V = \frac{s}{\overline{I}}$ \citep[where $s$ is the standard deviation of the flux density measurements and $\overline{I}$ is the average flux density of the source, as defined in ][]{swinbank2015}. These variability parameters are calculated for each unique source detected in the images. As we handle newly detected sources separately (see Section \ref{sec:transient_hunt}), in this section we only consider sources detected in the first observation. We process each observing frequency separately using {\sc TraP}, as the data quality between the observing frequency differs and this can impact the observed variability. In Figure \ref{fig:VandEta}, we plot the variability parameters for each of the sources at the four observing frequencies. We can model the distribution of stable sources as having a Gaussian distribution in both $\eta$ and $V$, with outliers being candidate variable sources \citep[][]{rowlinson2019,swinbank2015}. We define variable candidates as being $>1.5\sigma$ offset from the $\eta$ and $V$ distributions. In Table \ref{table:variability_limits}, we provide the $1.5\sigma$ thresholds used for each observing frequency and they are shown as the dashed black lines in Figure \ref{fig:VandEta}. This leads to 4 sources being identified as variable sources, including the XRB MAXI J1820+070 at the centre of the image as expected (highlighted with the red circle in Figure \ref{fig:VandEta}). The variable candidates are listed in the bottom part of Table \ref{table:variable_candidates} with their variability parameters for each observing frequency. Variability is confirmed by visual inspection and comparison to nearby sources. The multi-wavelength light curves of these sources are shown in Figures \ref{fig:lightcurves1} -- \ref{fig:lightcurves3}. Finally, we searched for associations within the NVSS 1.4 GHz radio catalogue \citep{condon1998}. Each of these variable sources are associated with known sources and their associations are given in Table \ref{table:variable_candidates}.

In addition to the variable sources identified by the {\sc TraP} in this analysis, we also consider the variable candidates found in the transient search conducted in Section \ref{sec:transient_hunt} given in the top part of Table \ref{table:variable_candidates}. We plot their variability parameters on Figure \ref{fig:VandEta} using red '+' symbols and note that they do not pass the variability threshold on $\eta$ as outlined in Table \ref{table:variability_limits}. This is likely due to the majority of flux density measurements being non-detections and hence the $\eta$ parameter is instead being dominated by the low variability in the noise in the images. 

In summary, we have identified three variable sources in addition to the target of these observations, MAXI J1820+070 \citep[which is considered in a separate publication;][]{bright2020}. In the following section, we will discuss these sources in more depth.

\section{Analysis of new variable sources}
\label{sec:analysis}

In this paper, we have identified three new variable sources via the transient search outlined in Section \ref{sec:transient_hunt}, and three new variable sources via the standard variability search given in Section \ref{sec:variable_hunt}. Here, we consider their multi-wavelength properties to determine the likely progenitor source. We use key diagnostic plots from \cite{stewart2018}\footnote{https://github.com/4pisky/radio-optical-transients-plot} and \cite{pietka2015}\footnote{https://github.com/FRBs/Transient\_Phase\_Space} to compare the properties of our sources to the known populations of transient and variable sources.

%\subsection{Analysis of transient sources}
%\label{sec:transient}

\subsection{PSR J1822+0705}

\begin{figure*}
\centering
\includegraphics[width=0.9\textwidth]{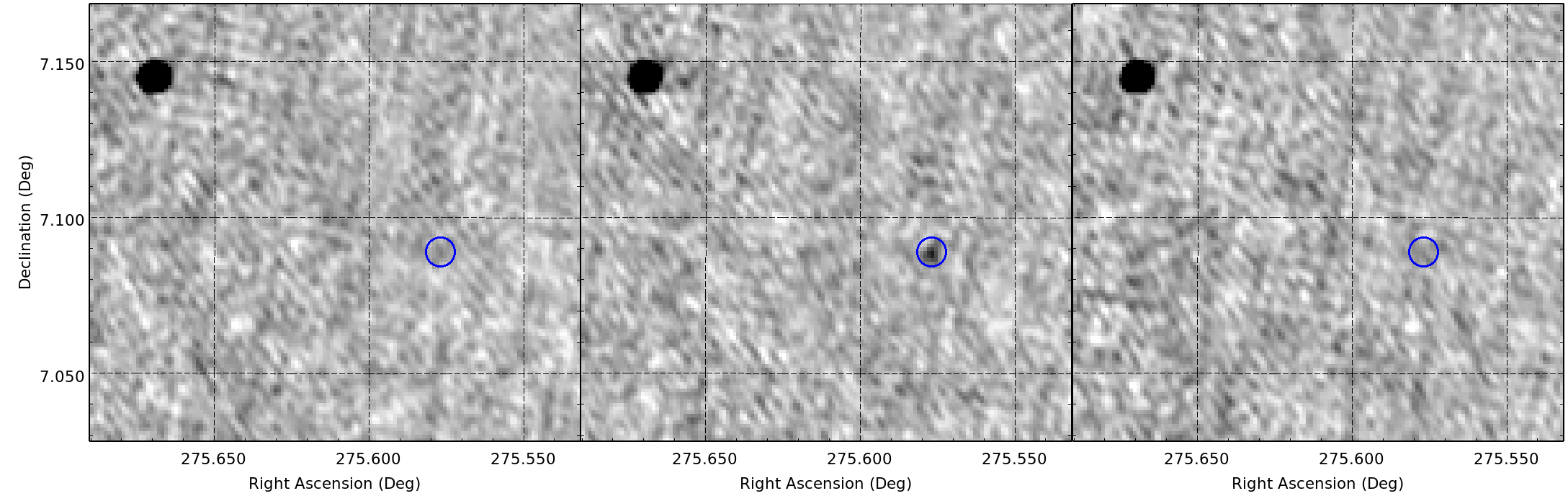}
\caption{Three 8 second snapshot images, with matching colour scales, of the location of PSR J1822+0705, consecutive in time, clearly showing the pulsar turning on and off during the observation.}
\label{fig:pulsarImg}
\end{figure*}

We searched the existing multi-wavelength catalogues by querying the location in {\sc Simbad} \citep{wenger2000}, and it is associated with PSR J1822+0705. PSR J1822+0705 is a regular pulsar with a spin period of 1.36 seconds, a flux density of 3.8$\pm$0.1 mJy at 400 MHz and a dispersion measure of 62.2 pc cm$^{-3}$ \citep[from the ATNF pulsar catalogue; ][]{manchester2005}\footnote{http://www.atnf.csiro.au/research/pulsar/psrcat}. This pulsar has not been extensively studied and was not previously known to be variable.

Pulsars are known to vary on short timescales, due to intrinsic effects, and on long timescales, due to both interstellar scintillation and intrinsic effects. To check if the variation observed is also seen on short timescales, we take the observation that recorded the brightest flux density for this pulsar (observation id 1543743065 on 2018-12-02 at 0.96 GHz) and re-imaged on the shortest timescales of 8 seconds, producing 112 snapshot images. In Figure \ref{fig:pulsarImg}, we show 3 consecutive 8 second images showing that the variation appears to be present on the shortest timescales we can image. The timescale of this variation is much shorter than both diffractive interstellar scintillation \citep[minutes to hours; ][]{hewish1968} and refractive interstellar scintillation \citep[days to years; ][]{sieber1982}. Therefore, this observed variation is likely to be intrinsic to the pulsar.

\begin{figure*}
\centering
\includegraphics[width=0.9\textwidth]{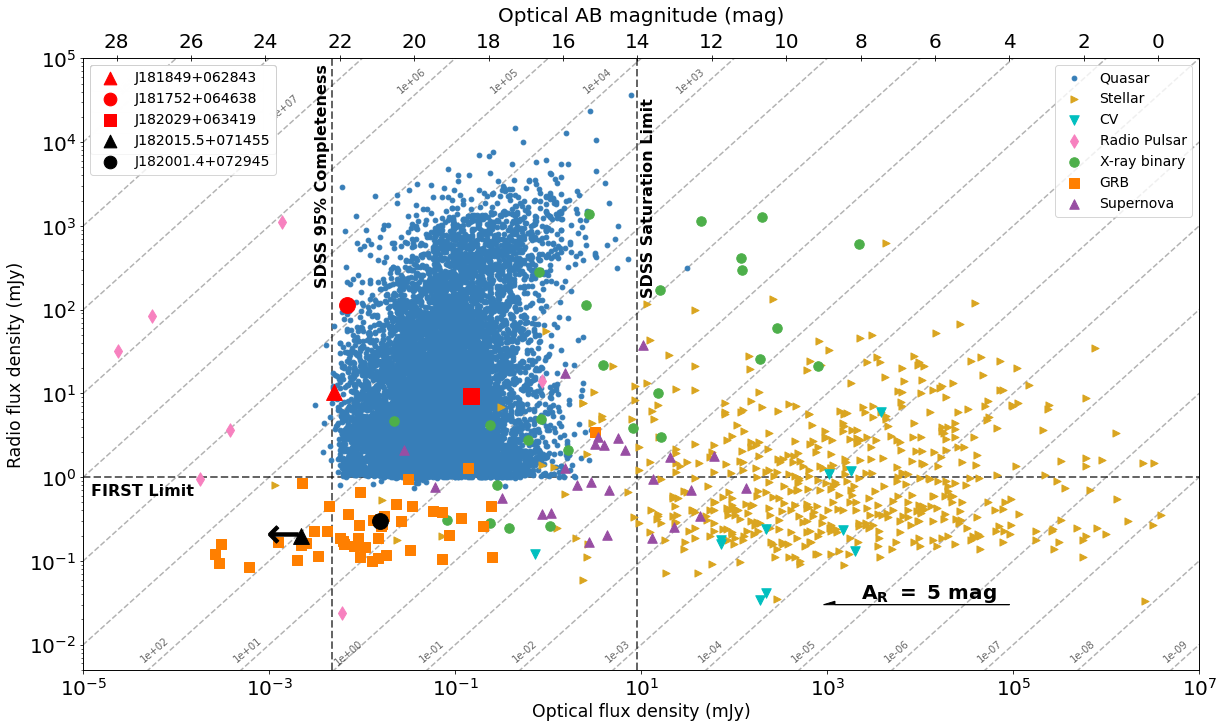}
\caption{Radio flux density versus optical flux density for different populations of transient and variable sources, adapted from Figure 1 in \citet{stewart2018}. The two unidentified variable sources identified in Section \ref{sec:transient_hunt} are shown with black symbols. The three variable sources identified in Section \ref{sec:variable_hunt} are shown with red symbols and are consistent with quasars.}
\label{fig:StewartFig}
\end{figure*}

\begin{figure*}
\centering
\includegraphics[width=0.9\textwidth]{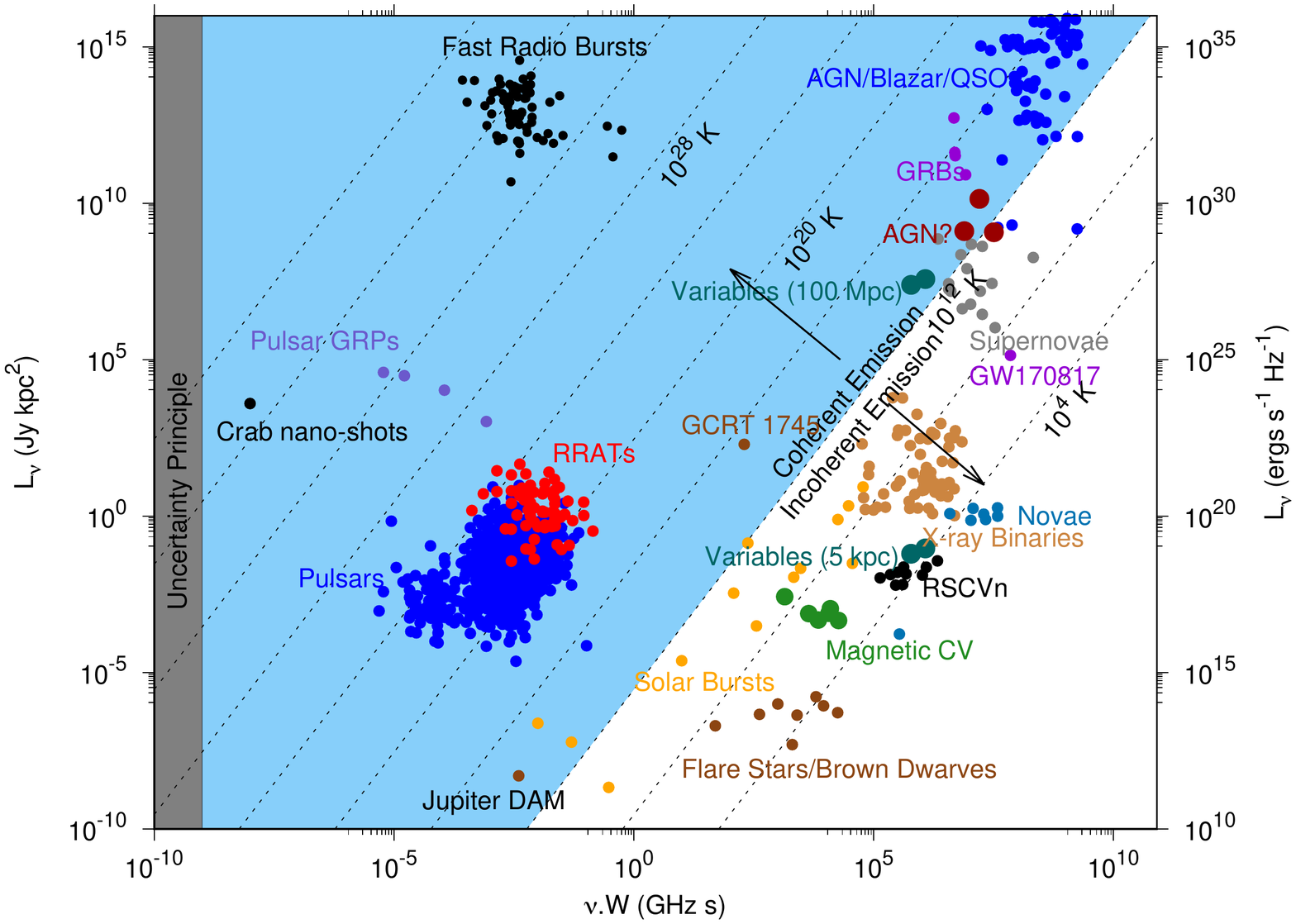}
\caption{Radio luminosity of different transient and variable sources as a function of their duration and observing frequency, adapted from \citet{pietka2015}. The diagonal lines show constant brightness temperature, with the blue and white regions showing coherent and incoherent sources, respectively. The two unidentified transient candidates (Section \ref{sec:transient_hunt}) are plotted in dark cyan, assuming they are Galactic (at 5 kpc) or extra-galactic (at 100 Mpc). In maroon, we plot the variable candidates, hypothesised to be AGNs, presented in Section \ref{sec:variable_hunt} assuming a distance of 100 Mpc.}
\label{fig:PietkaFig}
\end{figure*}

\subsection{MKT J182015.5+071455}
\label{sec:4.2}

This newly identified source is faint, with a maximum flux density of 0.2$\pm$0.1 mJy (at 1.39 GHz). It is visible in the deep images of this field, with a flux density of 0.089$\pm$0.013 mJy (at 1.39 GHz), showing that the source is likely persistent and has increased in brightness by a factor of $\sim$2 at its maximum. At this flux density, it is nearly an order of magnitude fainter than typically assumed limits for radio surveys like NVSS. There are no optical sources associated with this source in the PanSTARRS archive \citep{flewelling2020}, with a lower limit of 22.7 mag in the $i$ band. The deep radio image and optical image are shown in Figure \ref{fig:trans_25568}. This source is estimated to have a variability timescale of $\sim$1 weeks as it was typically observed in just one epoch with non-detections in the previous and following epochs. 

From Figure \ref{fig:StewartFig}, we note this source lies in a region of the plot where the optical survey used as a reference for AGN (SDSS) is not complete. While it is consistent with GRB afterglows in Figure \ref{fig:StewartFig}, this is unlikely given that the source appears to be persistent. Although poorly sampled, there are pulsars and stellar sources in this region. While there is potential of optical extinction towards the source, we note that the Galactic colour excess towards the source is E($g$-$r$) = 0.25 mag, suggesting $A_I\lesssim 0.5$ for Galactic sources \citep{green2019}. The most likely conclusion is that this source is consistent with a fainter AGN population than was sampled by SDSS.

In Figure \ref{fig:PietkaFig}, we consider the variability timescale of $\sim$2 weeks in comparison to other transient and variable populations. As the distance to this source is unknown, we take two typical examples: a Galactic origin with a distance of 5~kpc, and an extra-galactic origin with a distance 100 Mpc. If this source is Galactic, we determine that it is most likely an RSCVn or an X-ray binary; however, the typical optical luminosities of both classes argue against this conclusion. \cite{toet2021} have detected RS CVn at 144 MHz with flux densities of $\sim 0.5$ mJy), but these sources were only at 15--300 pc (with the sources above 100 pc always having at least one subgiant or giant star). At 300 pc, even low-mass dwarf stars should have been detected by PanSTARRs. In Figure 6, the optical fluxes of typical X-ray binaries in outburst were 100 times brighter than the optical limit. While we cannot rule out quiescent accretion onto an X-ray binary, only the quiescent radio flux of the black hole X-ray binary V404 Cyg \citep[$\sim2.4$ kpc;][]{millerjones2009} has been observed at $\sim 0.1$ mJy \citep[e.g.,][]{rana2016}. If it is extra-galactic, it may be consistent with a supernova, however the persistent nature of this source discounts this possibility, so it would then most likely be an AGN. For distances beyond the assumped 100 Mpc, the source becomes more consistent with the faint end of AGN, albeit with a shorter variability timescale. We note that the latter can be compressed by relativistic effects, but that is usually more severe for the brighter blazars. Although we cannot confidently conclude which type of astrophysical variable this source is, the leading contenders are  a pulsar, an atypical quiescently accreting X-ray binary, or an AGN  at a distance beyond 100 Mpc with a faster than average variability timescale. Given that a large fraction of variable radio sources are AGN, the latter may be the most likely of these conclusions.

\subsection{MKT J182001.4+072945}

As with the previous source, this newly detected source is faint with a flux density of 0.3$\pm$0.1 mJy at 1.39 GHz. This source is detected in the deep image with a flux density of 0.13$\pm$0.02 mJy (1.39 GHz), showing that the source is likely persistent and has increased in brightness by a factor of $\sim$3 at its maximum. We find this source is associated with the PanSTARRS source PSO J275.0057+07.4958, which is detected in all observing filters except the $g$ band. The observed magnitudes of this optical sources are $r$=21.26$\pm$0.07, $i$=20.58$\pm$0.05, $z$=20.27$\pm$0.08, and $y$=19.87$\pm$0.05 \citep{flewelling2020}. The deep radio image and optical image are shown in Figure \ref{fig:trans_20764}. This source is estimated to have a variability timescale of $\sim$2 weeks as it was typically observed in two consecutive epochs with non-detections in the previous and following epochs.

The maximum radio flux density of 0.5 mJy at 0.96 GHz and variability timescale of $\sim2$ weeks is similar to  MKT J182015.5+071455 above. Thus the conclusions discussed in Section \ref{sec:4.2} based just on radio flux density and timescale are largely the same. However, with an optical brightness of $\sim$20 mag, some of the Galactic scenarios dismissed need to be reevaluated. At distances of 300 pc and 5 kpc, the absolute optical magnitude is roughly $\sim$12 mag and $\sim$6.5 mag, respectively. The former still tends to rule out most dwarf stars and thus the RS CVn category. However, the latter is consistent with a late K dwarf and suggests that a somewhat nearby X-ray binary, particularly one in quiescence, cannot be ruled out. Although we cannot confidently conclude which type of astrophysical variable this source is, the leading contenders are  a pulsar, an X-ray binary, or an AGN  at a distance beyond 100 Mpc with a faster than average variability timescale. Again, an AGN is the most likely of these conclusions given the typical source densities of such sources.

%\subsection{Analysis of variable sources}
%\label{sec:variable}

\subsection{NVSS J181849+062843}

First, we searched for associations within existing radio catalogues at 1.4 GHz, specifically NVSS \citep{condon1998} and VLASS \citep{gordon2021}. We find that this source is associated with NVSS J181849+062843, and has flux densities of 2.2$\pm$0.4 mJy (1.4 GHz; NVSS) and 13.8$\pm$0.3 mJy (2--4 GHz; VLASS), respectively. We note that the average flux density at 1.39 GHz in these observations is five times higher than the observed NVSS flux density, confirming this source is variable on long timescales. 

We searched the existing multi-wavelength catalogues by querying the location in {\sc Simbad} \citep{wenger2000} and no associated sources are found. To search for a fainter optical source association, we queried the PanSTARRS catalogue \citep{flewelling2020}. We find that NVSS J181849+062843 is associated with the PanSTARRS source PSO J274.7073+06.4791, which was detected in just two of the PanSTARRS observing filters: $i$=21.80$\pm$0.20 mag and $z$=20.88$\pm$0.02 mag. In Figure \ref{fig:274.707}, we show the MeerKAT and PanSTARRS images of this source.

With an optical brightness of $\sim$21 mag and an average radio flux density of 10 mJy at 1.39 GHz, this source is most likely an AGN or quasar, as shown in Figure \ref{fig:StewartFig}. With a variability timescale of $\sim$1--3 months (determined by visual inspection of peaks in the light curve), this is typically very short for observed AGN intrinsic variability \citep[on timescales $>$days to years; e.g.,][]{hovatta2008} but not inconsistent as shown in Figure \ref{fig:PietkaFig}.
Thus, we conclude that this source is most likely an AGN.

\subsection{NVSS J181752+064638}

This bright variable source is associated with NVSS 181752+064638 with an NVSS flux density of 100.8$\pm$3.1 mJy at 1.4 GHz \citep{condon1998}. This source is also detected in the VLASS survey with a flux density of 73.0$\pm$0.3 mJy at 2--4 GHz \citep{gordon2021}. We find this source is on average 10\% brighter in our MeerKAT observations. The VLASS flux density at 2--4 GHz is $\sim$30\% fainter, signifying either spectral behavior or variability.

We searched the existing multi-wavelength catalogues by querying the location in {\sc Simbad} \citep{wenger2000}, and NVSS J181752+064638 has a flat spectrum and has been identified as a blazar \citep{healey2007}. Additionally, we find this source is associated with the PanSTARRS source PSO J274.4693+06.7773, which is detected in all observing filters except $g$. The observed magnitudes of the optical counterpart are $r$=21.80$\pm$0.02, $i$=21.45$\pm$0.03, $z$=21.09$\pm$0.21, and $y$=20.39$\pm$0.19 \citep{flewelling2020}. In Figure \ref{fig:274.469}, we show the MeerKAT and PanSTARRS images of this source.

With an optical brightness of $\sim$21 mag and an average radio flux density of 113 mJy at 1.39 GHz, this source is most likely an AGN, as shown in Figure \ref{fig:StewartFig}, further supporting the blazar identification. A variability timescale of $\sim$3--6 months (as determined by visual inspection of peaks in the light curve) is relatively short for an AGN but is not inconsistent as shown in Figure \ref{fig:PietkaFig}.
Thus, we conclude that this source is also most likely an AGN.
 
\subsection{NVSS J182029+063419}

This source was also associated with sources in NVSS (NVSS J182029+063419) and VLASS, with flux densities of 5.8$\pm$0.4 mJy (1.4 GHz) and 13.4$\pm$0.3 mJy (2--4 GHz), respectively \citep{gordon2021,condon1998}. Our MeerKAT observations show that the flux density has approximately doubled since the NVSS observations.

We searched the existing multi-wavelength catalogues by querying the location in {\sc Simbad} \citep{wenger2000} and no associated sources were found. We queried the PanSTARRS catalogue and find that the bright point source PSO J275.1254+06.5721 is associated. The optical source is detected in all of the PanSTARRS filters with observed magnitudes of $g$=18.89$\pm$0.02, $r$=18.27$\pm$0.01, $i$=18.18$\pm$0.01, $z$=17.91$\pm$0.02, and $y$=17.78$\pm$0.03 \citep{flewelling2020}. 

With an optical brightness of $\sim$18 mag and an average radio flux density of 13 mJy at 1.39 GHz, this source could be either a quasar or an XRB, as shown in Figure \ref{fig:StewartFig}. With an observed variability timescale of $\sim$1 year (determined by visual inspection of peaks in the light curve), this variable source could be an XRB, but is more likely an AGN. We show this source in Figure \ref{fig:PietkaFig}, assuming it is extra galactic at a distance of 100 Mpc. We note that, with a Galactic distance of 5 kpc, this source could still be associated with XRBs or novae. 

Therefore, we determine that this source is either an AGN, a nova or an XRB. Further radio observations or an optical spectrum will be able to confirm this identification.

\section{Discussion \& Conclusions}
\label{sec:conclusions}

In Section \ref{sec:transient_hunt}, we presented a transient hunt on 1 week timescales for 62 epochs and a search radius of 0.9 degrees, giving a total sky area surveyed of 155 square degrees to a limiting sensitivity of 1 mJy. A known pulsar was found via this search, and also two transient candidates that were reclassified as variables, as they were associated with persistent emission in deep images. The 1 week timescales for transient sources at $\sim$1.4 GHz is a relatively unexplored parameter space, as shown in the left hand panel of Figure \ref{fig:surfaceDensityPlot}, with only one other survey on comparable timescales by \cite{bell2011}. The transient surface density limit presented in this work is similar to that attained by \cite{bell2011} using 5037 epochs of observations from the Very Large Array; however, our survey is an order of magnitude more sensitive than that survey and using 2 orders of magnitude fewer observations. Thus, with a relatively small sample of MeerKAT data, we are able to place highly competitive limits on the presence of transients on timescales of $\sim$1 week. Significantly more observations on these timescales have been obtained by monitoring XRBs using MeerKAT; when processed, these will be able to determine if there is a rare population of transient sources on these timescales.

The known pulsar PSR J1822+0705 that was detected as part of the transient hunt, was previously poorly studied and was not known to show variable emission. We identified it as varying on weekly timescales and further investigation of the brightest epoch demonstrated it is varying on timescales of seconds. This short timescale variability indicates that it is intrinsic to the pulsar and is not caused by interstellar scintillation. From studies of known variable pulsars \citep[e.g.,][]{parent2022}, there are a number of possible causes for intrinsic variation including giant pulses \citep[e.g.,][]{lundgren1995}, nulling pulsars \citep[e.g.,][]{backer1970}, intermittent pulsars \citep[e.g.,][]{kramer2006}, and variable flux density emission states \citep[e.g.,][]{young2014}. Further detailed analysis is required to determine the origin of the intrinsic variability of PSR J1822+0705. Imaging surveys are an excellent method to detect extreme pulsar behaviour and a method to discover new highly variable pulsars. This is because large areas of sky can be monitored over long timescales, which can be challenging with traditional pulsar timing methods \citep[e.g.,][]{hobbs2016}.

In addition to PSR J1822+0705, we detected two other likely variable sources. Based on their variability timescales and radio luminosities, we determined that they are most likely AGNS though one source could also be a Galactic pulsar or a quiescent X-ray binary. Though it is unexpected that AGN vary on timescales $\sim$1--2 weeks, this may be subject to observational biases in AGN surveys. Alternatively, the observed variability timescale may only show the peak of the emission due to the sources being near to the detection threshold in this survey and the ``true'' variability timescale could be much longer.

In addition to a transient search, we also conducted a search for variable sources in the field in Section \ref{sec:variable_hunt}. We identified three likely AGNs/blazars that are varying on timescales from $\sim$1 month up to $\sim$1 year. Although not unexpected, the number of AGNs known to vary on these timescales is relatively small \citep[e.g.,][]{pietka2015}, though the known population is growing rapidly \citep{driessen2022b,sarbadhicary2021}. By identifying the variability of these sources via imaging surveys, we are able to significantly increase the known population of variable AGNs, especially on short timescales, leading to a greater understanding of these sources. Following identification, dedicated AGN studies can determine if the variability is an intrinsic effect caused by, for example, variable accretion rates on to the central black hole or if the variability is caused by extrinsic effects such as scintillation.

In summary, this paper has presented an optimised transient and variability search method for MeerKAT observations. The transient limits attained are competitive despite this dataset being comparatively small. Finally, we identified four AGN varying on relatively short timescales, one source that may be a pulsar or a quiescent X-ray binary (though, based on statistical arguments, also most likely to be an AGN) and one pulsar demonstrating extreme variability on both short (eight seconds) and long (weeks) timescales.

\section*{Acknowledgements}

We thank Ben Stappers for useful discussions regarding this work. AR acknowledges funding from the NWO Aspasia grant (number: 015.016.033). PAW acknowledges financial support from the University of Cape Town and the National Research Foundation. GRS is supported by NSERC Discovery Grants RGPIN-2016-06569 and RGPIN-2021-0400.

We acknowledge use of the Inter-University Institute for Data Intensive Astronomy (IDIA) data intensive research cloud for data processing. IDIA is a South African university partnership involving the University of Cape Town, the University of Pretoria and the University of the Western Cape. 

The MeerKAT telescope is operated by the South African Radio Astronomy Observatory (SARAO), which is a facility of the National Research Foundation, an agency of the Department of Science and Innovation. We would like to thank the operators, SARAO staff and ThunderKAT Large Survey Project team. 

This research made use of Astropy, a community-developed core Python package for Astronomy \citep{astropy, price2018}. We also use {\sc Numpy} \citep{harris2020}, {\sc Scipy} \citep{virtanen2020}, {\sc pandas} \citep{pandas}, {\sc SQLAlchemy}\footnote{https://www.sqlalchemy.org} and {\sc matplotlib} \citep{hunter2007}. This research has made use of the SIMBAD database, operated at CDS, Strasbourg, France.

The Pan-STARRS1 Surveys (PS1) and the PS1 public science archive have been made possible through contributions by the Institute for Astronomy, the University of Hawaii, the Pan-STARRS Project Office, the Max-Planck Society and its participating institutes, the Max Planck Institute for Astronomy, Heidelberg and the Max Planck Institute for Extraterrestrial Physics, Garching, The Johns Hopkins University, Durham University, the University of Edinburgh, the Queen's University Belfast, the Harvard-Smithsonian Center for Astrophysics, the Las Cumbres Observatory Global Telescope Network Incorporated, the National Central University of Taiwan, the Space Telescope Science Institute, the National Aeronautics and Space Administration under Grant No. NNX08AR22G issued through the Planetary Science Division of the NASA Science Mission Directorate, the National Science Foundation Grant No. AST-1238877, the University of Maryland, Eotvos Lorand University (ELTE), the Los Alamos National Laboratory, and the Gordon and Betty Moore Foundation.

%%%%%%%%%%%%%%%%%%%%%%%%%%%%%%%%%%%%%%%%%%%%%%%%%%
\section*{Data Availability}

The data and scripts underlying this article are available in Zenodo, at \url{https://zenodo.org/record/6826588#.Ys60VC8Rpqs}. The MeerKAT data used are available in the MeerKAT data archive at \url{https://archive.sarao.ac.za/}. The PanSTARRS fits images were obtained from \url{https://panstarrs.stsci.edu}.



\bibliographystyle{mnras}
\bibliography{MeerKAT_Variable.bib} % if your bibtex file is called example.bib

%\begin{thebibliography}{99}
%\end{thebibliography}

%%%%%%%%%%%%%%%%%%%%%%%%%%%%%%%%%%%%%%%%%%%%%%%%%%

%%%%%%%%%%%%%%%%% APPENDICES %%%%%%%%%%%%%%%%%%%%%

%\appendix



%%%%%%%%%%%%%%%%%%%%%%%%%%%%%%%%%%%%%%%%%%%%%%%%%%

\appendix
\section{Observations}

\begin{table*}
\centering
\begin{tabular}{|c c c l|} 
\hline
Number & Observation Date & Observation ID & Rejected \\
 & (UTC) &  &  \\
\hline
1  & 28/09/2018 17:43 & 1538156623 & Failed automated processing \\
2  & 05/10/2018 16:30 & 1538757039 & Failed automated processing \\
3  & 11/10/2018 16:04 & 1539253756 & Failed automated processing \\
4  & 12/10/2018 19:31 & 1539354654 & Failed automated processing \\
5  & 14/10/2018 15:01 & 1539529257 & Failed automated processing \\
6  & 19/10/2018 14:44 & 1539955889 & rms high for all frequencies \\
7  & 27/10/2018 12:49 & 1540640133 &  \\
8  & 03/11/2018 11:54 & 1541242861 & rms high for all frequencies \\
9  & 10/11/2018 10:33 & 1541845876 & Failed automated processing \\
10 & 13/11/2018 15:31 & 1542123060 & Failed automated processing \\
11 & 17/11/2018 11:26 & 1542451956 &  \\
12 & 24/11/2018 10:39 & 1543053930 & rms high for 1.39 GHz \\
13 & 02/12/2018 10:05 & 1543743065 & rms high for 1.39 GHz \\
14 & 08/12/2018 09:38 & 1544259897 & rms high for all frequencies \\
15 & 15/12/2018 14:05 & 1544882682 & Failed automated processing \\
16 & 22/12/2018 13:52 & 1545484573 &  \\
17 & 29/12/2018 13:21 & 1546087573 &  \\
18 & 05/01/2019 11:51 & 1546686970 &  \\
19 & 12/01/2019 11:11 & 1547288173 & rms high for 0.96 GHz and 1.39 GHz \\
20 & 19/01/2019 08:59 & 1547884999 & Failed automated processing \\
21 & 26/01/2019 08:59 & 1548489789 &  \\
22 & 01/02/2019 04:45 & 1548991875 &  \\
23 & 09/02/2019 04:55 & 1549688122 & Failed automated processing \\
24 & 09/03/2019 01:36 & 1552093390 & rms high for 0.96 GHz and 1.39 GHz \\
25 & 18/03/2019 03:15 & 1552872672 & rms high for 0.96 GHz and 1.39 GHz \\
26 & 25/03/2019 03:14 & 1553477469 & rms high for 0.96 GHz and 1.39 GHz \\
27 & 01/04/2019 02:54 & 1554080984 & rms high for 0.96 GHz and 1.39 GHz \\
28 & 09/04/2019 02:45 & 1554771678 & rms high for 0.96 GHz and 1.39 GHz \\
29 & 15/04/2019 02:46 & 1555290074 & rms high for 1.39 GHz \\
30 & 20/04/2019 03:17 & 1555723938 & rms high for 0.96 GHz and 1.39 GHz \\
31 & 29/04/2019 04:58 & 1556507556 &  \\
32 & 04/05/2019 23:15 & 1557005454 &  \\
33 & 11/05/2019 22:56 & 1557610250 &  \\
34 & 18/05/2019 22:41 & 1558213252 & rms high for 1.39 GHz \\
35 & 25/05/2019 01:37 & 1558743128 &  \\
36 & 10/08/2019 18:48 & 1565459161 &  \\
37 & 16/08/2019 21:14 & 1565987538 & rms high for 0.96 GHz \\
38 & 23/08/2019 16:13 & 1566574272 &  \\
39 & 31/08/2019 18:14 & 1567272657 &  \\
40 & 07/09/2019 16:07 & 1567871158 &  \\
41 & 14/09/2019 17:59 & 1568482488 &  \\
42 & 21/09/2019 14:55 & 1569076254 &  \\
43 & 29/09/2019 15:31 & 1569768359 &  \\
44 & 06/10/2019 15:56 & 1570375895 &  \\
45 & 12/10/2019 17:50 & 1570901089 & rms high for 0.96 GHz and 1.39 GHz \\
46 & 19/10/2019 15:01 & 1571494725 & rms high for 1.39 GHz \\
47 & 26/10/2019 14:52 & 1572097857 &  \\
48 & 01/11/2019 16:49 & 1572623454 &  \\
49 & 21/02/2020 06:48 & 1582262704 &  \\
50 & 02/03/2020 04:18 & 1583118973 &  \\
51 & 09/03/2020 02:54 & 1583721066 &  \\
52 & 14/03/2020 08:09 & 1584171959 &  \\
53 & 21/03/2020 01:29 & 1584753662 &  \\
54 & 29/03/2020 18:04 & 1585466272 & Failed automated processing \\
55 & 03/04/2020 03:16 & 1585883785 & Failed automated processing \\
56 & 10/04/2020 07:04 & 1586498460 & rms high for 1.39 GHz \\
57 & 18/04/2020 03:06 & 1587176616 &  \\
58 & 25/04/2020 04:29 & 1587786356 &  \\
59 & 02/05/2020 00:08 & 1588375604 &  \\
60 & 09/05/2020 03:49 & 1588992355 &  \\
61 & 17/05/2020 03:49 & 1589683557 &  \\
62 & 22/05/2020 23:21 & 1590189641 & Failed automated processing \\
\hline
\end{tabular}
\caption{The start time of each observation attained of the field of MAXI J1820+070.}
\label{table:observations}
\end{table*}

\begin{table*}
\centering
\begin{tabular}{|c c c l|} 
\hline
Number & Observation Date & Observation ID & Rejected \\
 & (UTC) &  &  \\
\hline
63 & 01/06/2020 02:01 & 1590975056 &  \\
64 & 08/06/2020 03:09 & 1591585260 &  \\
65 & 14/06/2020 21:04 & 1592167259 &  \\
66 & 19/06/2020 01:04 & 1592527255 &  \\
67 & 26/06/2020 23:51 & 1593213962 &  \\
68 & 04/07/2020 19:57 & 1593890159 &  \\
69 & 12/07/2020 19:43 & 1594580523 &  \\
70 & 19/07/2020 19:48 & 1595185558 &  \\
71 & 24/07/2020 21:35 & 1595623962 &  \\
72 & 01/08/2020 17:09 & 1596300367 &  \\
73 & 08/08/2020 16:54 & 1596904261 &  \\
74 & 15/08/2020 18:40 & 1597515368 &  \\
75 & 22/08/2020 21:24 & 1598130064 &  \\
76 & 30/08/2020 16:20 & 1598801721 &  \\
77 & 04/09/2020 20:55 & 1599251463 &  \\
\hline
\end{tabular}
%\caption{The start time of each observation attained of the field of MAXI J1820+70.}
\contcaption{}
\end{table*}

\section{Previous transient surveys}

\begin{table*}
\centering
\begin{tabular}{|c c c l|} 
\hline
Sensitivity & Transient Surface Density & Timescale & Author \\
(Jy) & (deg$^{-2}$) & (days) & \\
\hline
$6 \times 10^{-3}$   & $<0.14$                & $365$                & \cite{riley1998} \\ 
$6 \times 10^{-3}$   & $4.4 \times 10^{-4}$   & $365$                & \cite{levinson2002} \\
$1.5 \times 10^{-4}$ & $<5.7$                 & $19$                 & \cite{carilli2003} \\
$2 \times 10^{-3}$   & $< 1.6 \times 10^{-2}$ & $2555$               & \cite{devries2004} \\
$1 \times 10^{-2}$   & $<0.3$                 & $30$                 & \cite{bower2010} \\
$1 \times 10^{-3}$   & $<1$                   & $30$                 & \cite{bower2010} \\
$4 \times 10^{-2}$   & $<4 \times 10^{-3}$    & $81$                 & \cite{croft2010} \\
$7 \times 10^{-2}$   & $<3 \times 10^{-3}$    & $1$                  & \cite{bower2011} \\
$3$                  & $<9 \times 10^{-4}$    & $1$                  & \cite{bower2011} \\
$0.35$               & $<6 \times 10^{-4}$    & $1$                  & \cite{croft2011} \\
$8 \times 10^{-3}$   & $<3.2 \times 10^{-2}$  & $4$                  & \cite{bell2011} \\
$1 \times 10^{-3}$   & $6.5 \times 10^{-3}$   & $2 \times 10^{-3}$   & \cite{thyagarajan2011} \\
$1 \times 10^{-3}$   & $6.5 \times 10^{-3}$   & $365$                & \cite{thyagarajan2011} \\
$3.7 \times 10^{-4}$ & $<0.6$                 & $1.4 \times 10^{-2}$ & \cite{frail2012} \\
$2 \times 10^{-4}$   & $<3$                   & $60$                 & \cite{frail2012} \\
$9 \times 10^{-5}$   & $<6$                   & $365$                & \cite{frail2012} \\
$1 \times 10^{-3}$   & $<5 \times 10^{-2}$    & $2555$               & \cite{hodge2013} \\
$1 \times 10^{-3}$   & $<5 \times 10^{-2}$    & $5110$               & \cite{hodge2013} \\
$2.1 \times 10^{-4}$ & $<0.37$                & $1$                  & \cite{mooley2013} \\
$3$                  & $2 \times 10^{-6}$     & $1$                  & \cite{aoki2014} \\
$5 \times 10^{-4}$   & $<17$                  & $2 \times 10^{-2}$   & \cite{alexander2015} \\
$6.9 \times 10^{-5}$ & $<7.5$                 & $75$                 & \cite{bell2015} \\
$6 \times 10^{-4}$   & $<8 \times 10^{-2}$    & $1$                  & \cite{mooley2016} \\
$1 \times 10^{-3}$   & $<0.1$                 & $182.5$              & \cite{hancock2016} \\
$1 \times 10^{-3}$   & $<0.1$                 & $2555$               & \cite{hancock2016} \\
$1.5 \times 10^{-3}$ & $<0.3$                 & $1$                  & \cite{bhandari2018} \\
$1.5 \times 10^{-3}$ & $<0.3$                 & $12$                 & \cite{bhandari2018} \\
$1.4 \times 10^{-2}$ & $<3 \times 10^{-2}$    & $4745$               & \cite{bhandari2018} \\ 
$5.4 \times 10^{-5}$ & $<5.2$                 & $2555$               & \cite{radcliffe2019} \\
$5.4 \times 10^{-5}$ & $<5.2$                 & $8030$               & \cite{radcliffe2019} \\
$1.2 \times 10^{-3}$ & $4.9 \times 10^{-3}$   & $30$                 & \cite{murphy2021} \\
$1 \times 10^{-3}$   & $<3.7 \times 10^{-2}$  & $7$                  & This Work \\

\hline
\end{tabular}
\caption{The transient surface density constraints at $\sim$1.4 GHz obtained using a range of transient surveys. These data are used to plot Figure \ref{fig:surfaceDensityPlot} and were obtained from \citet{mooley2016} \tablefootnote{ \url{http://www.tauceti.caltech.edu/kunal/radio-transient-surveys/index.html}}.} \label{table:transientSDs}
\end{table*}

\section{Radio and optical images of variable sources}

\begin{figure}
\centering
\includegraphics[width=0.48\textwidth]{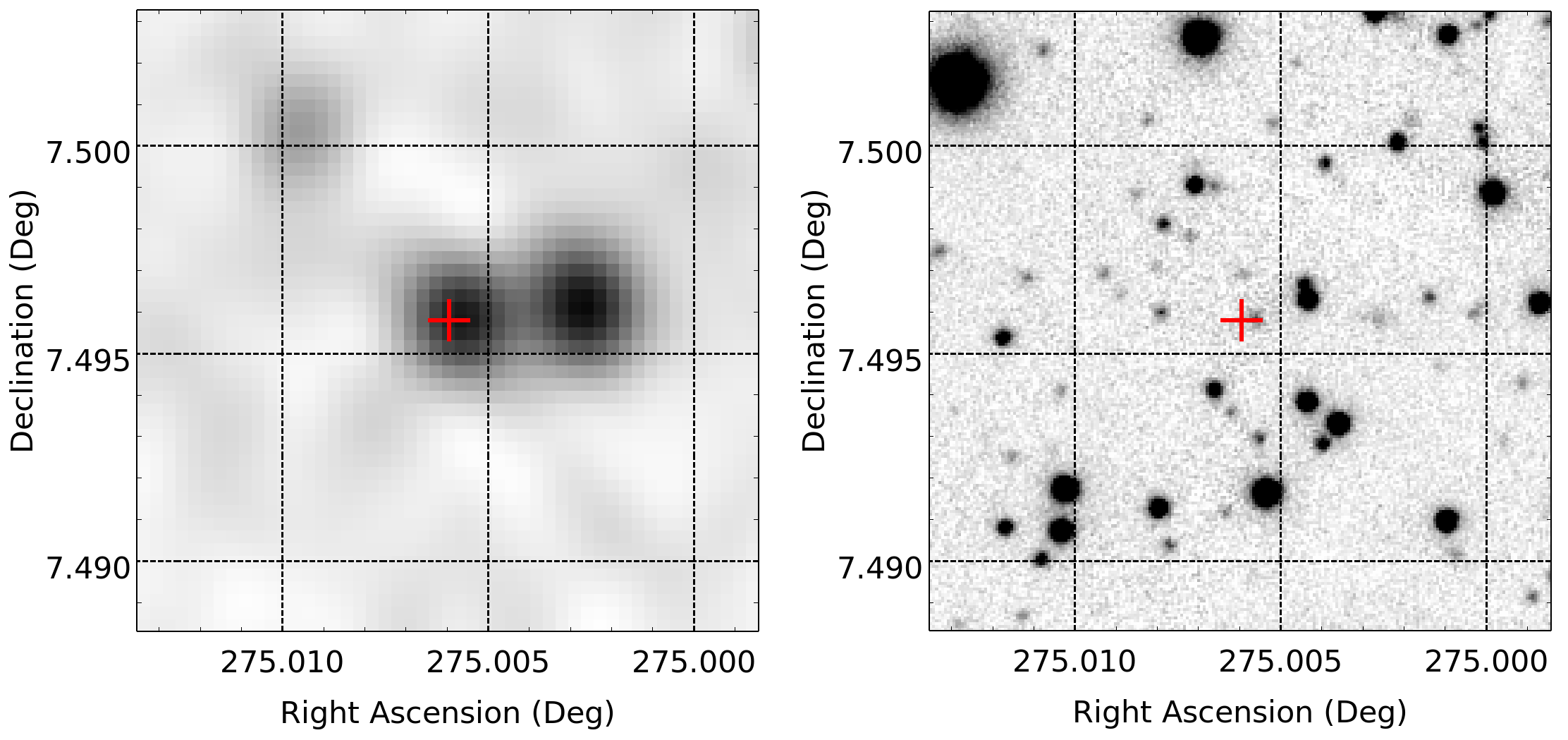}
\caption{MKT J182015.5+071455. Left: deep MeerKAT image. Right: PanSTARRS $z$ band image. The red plus symbol shows the location of the source.}
\label{fig:trans_25568}
\end{figure}

\begin{figure}
\centering
\includegraphics[width=0.48\textwidth]{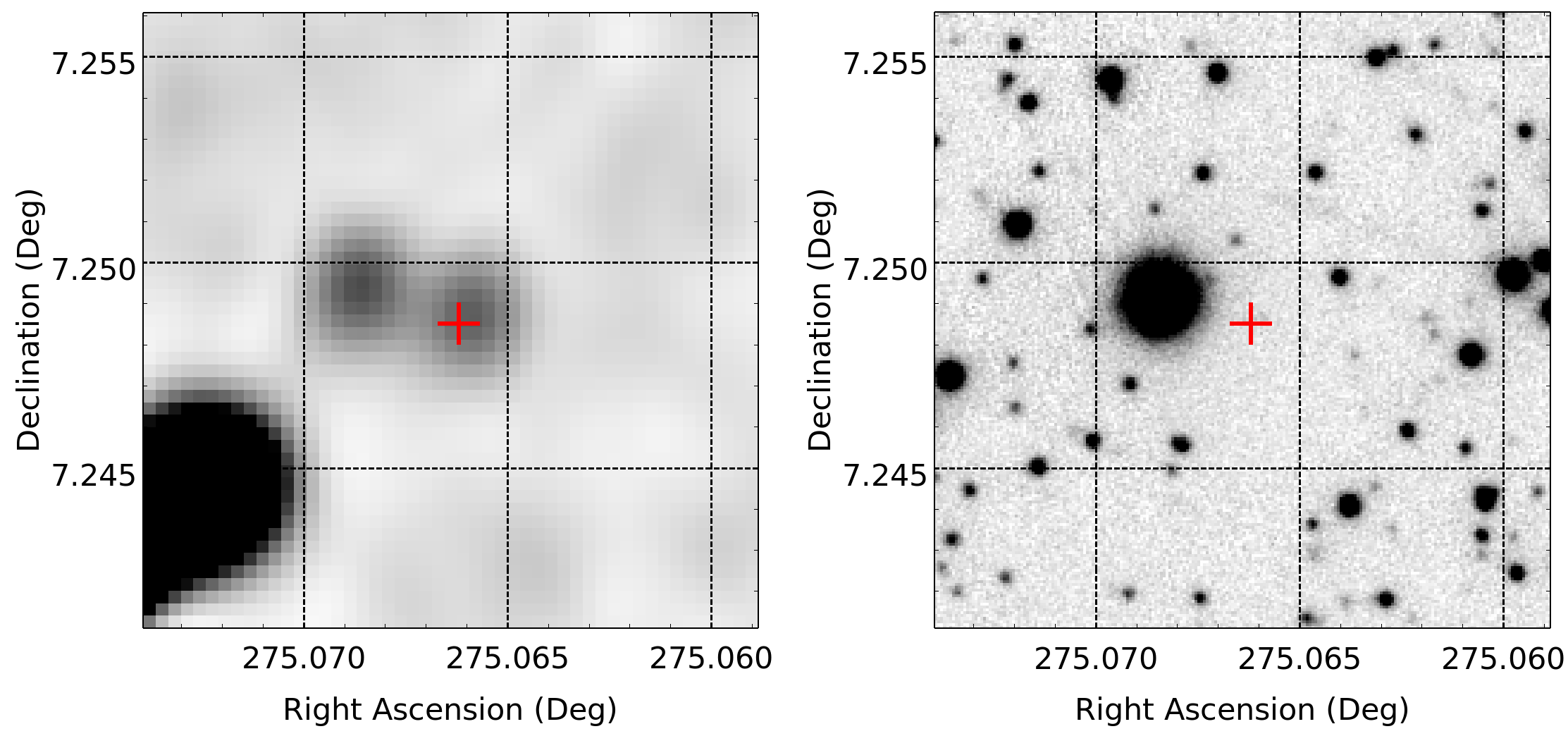}
\caption{MKT J182001.4+072945. Left: deep MeerKAT image. Right: PanSTARRS $z$ band image. The red plus symbol shows the location of the source.}
\label{fig:trans_20764}
\end{figure}

\begin{figure}
\centering
\includegraphics[width=0.45\textwidth]{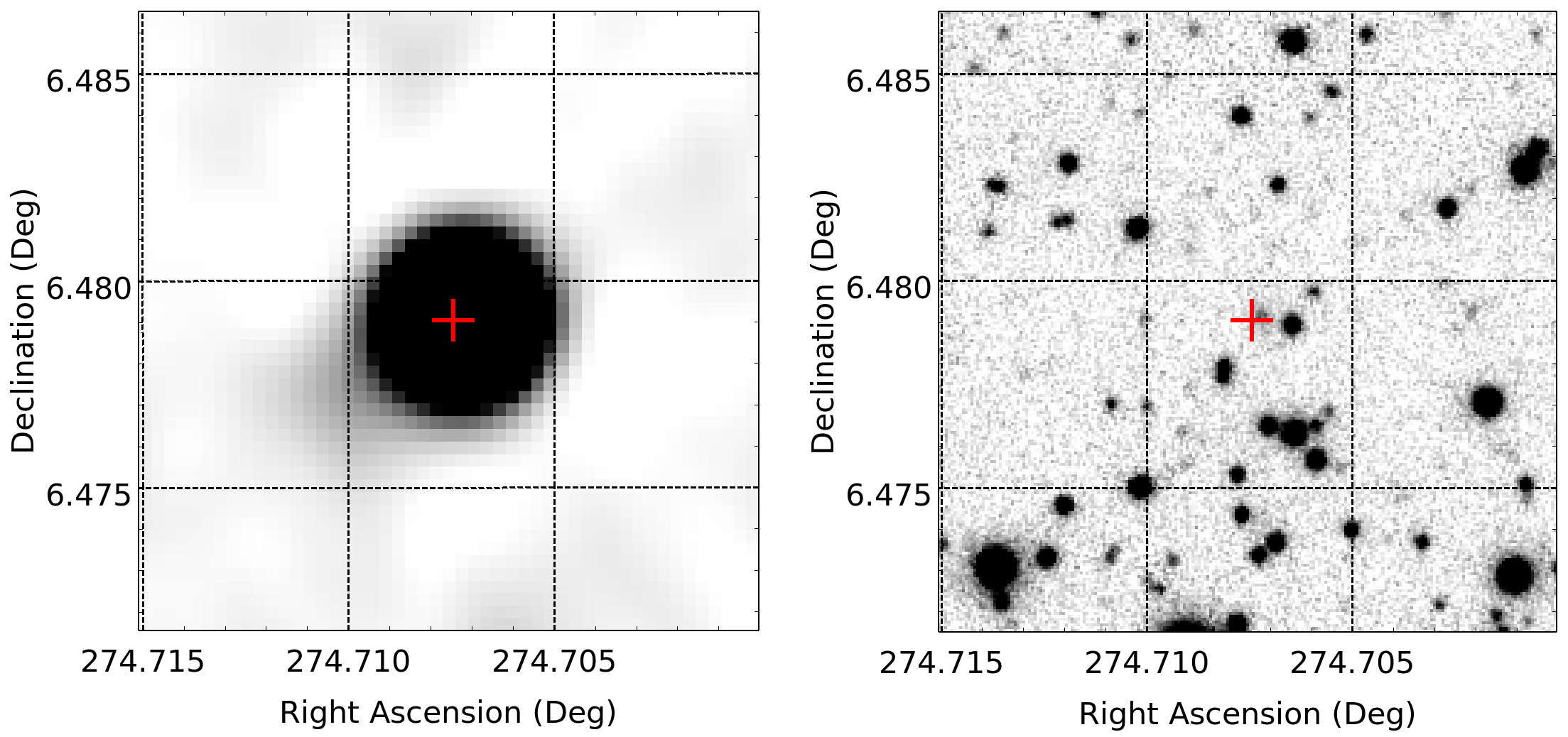}
\caption{NVSS J181849+062843. Left: deep MeerKAT image. Right: PanSTARRS $z$ band image. The red plus symbol shows the location of the source.}
\label{fig:274.707}
\end{figure}

\begin{figure}
\centering
\includegraphics[width=0.45\textwidth]{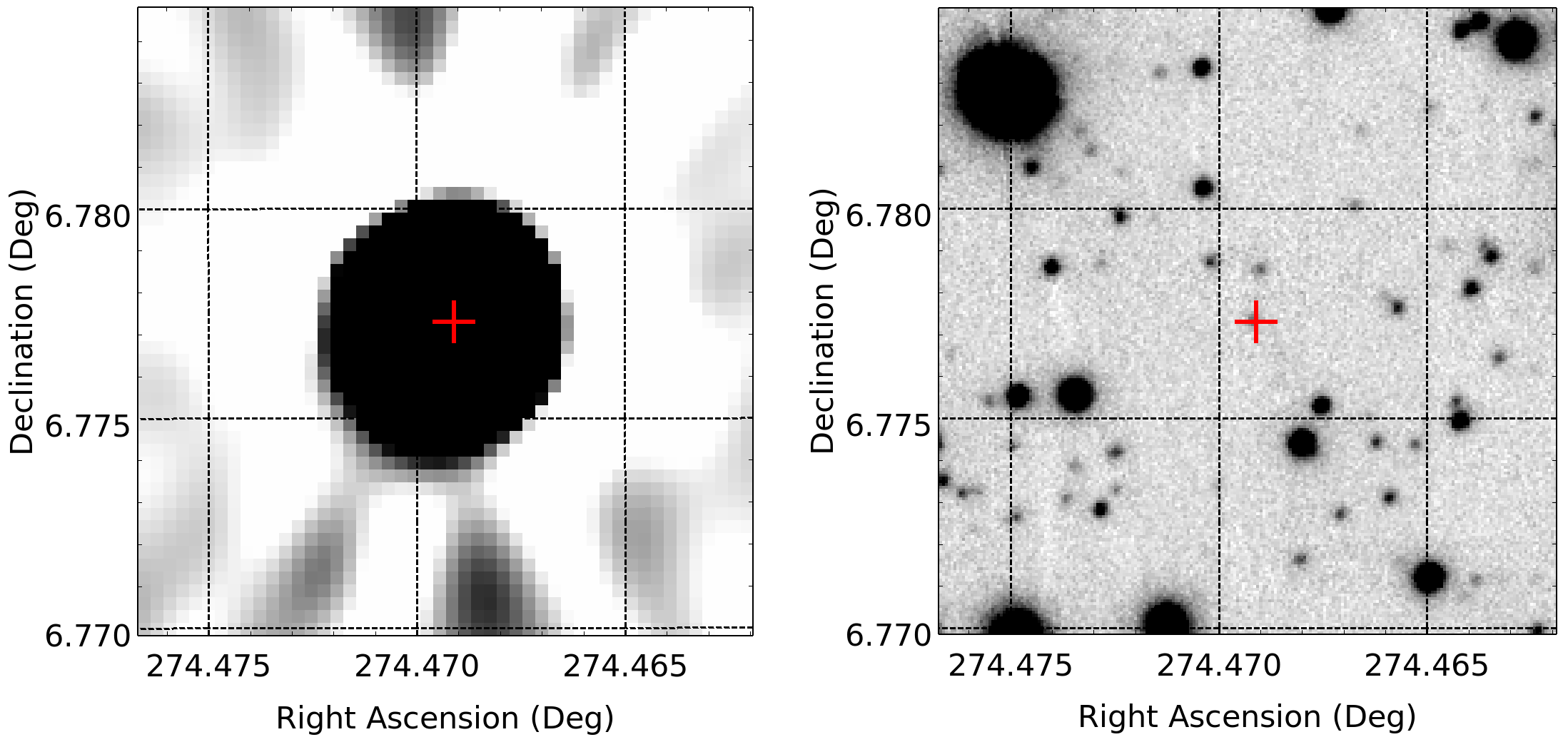}
\caption{NVSS J181752+064638. Left: deep MeerKAT image. Right: PanSTARRS $z$ band image. The red plus symbol shows the location of the source.}
\label{fig:274.469}
\end{figure}

\begin{figure}
\centering
\includegraphics[width=0.45\textwidth]{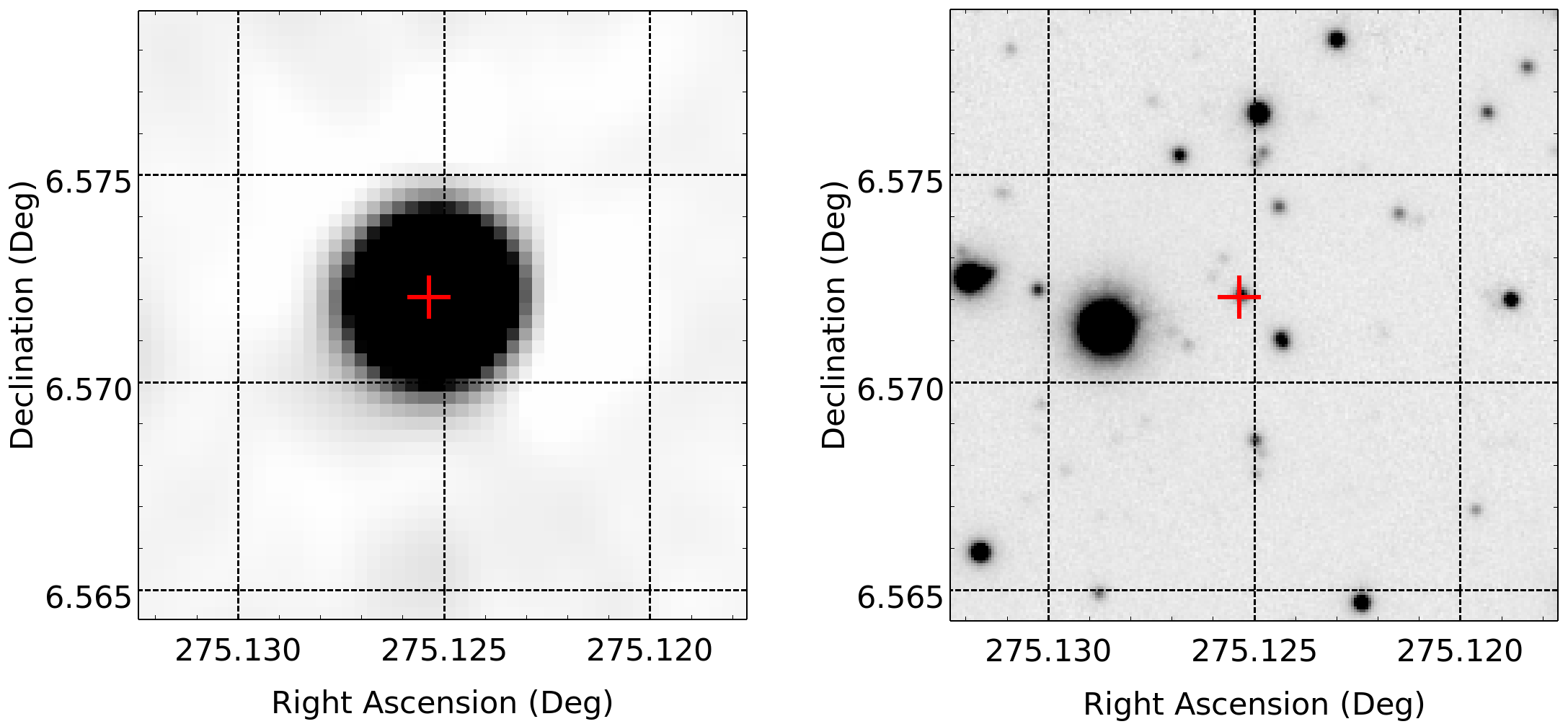}
\caption{NVSS J182029+063419. Left: deep MeerKAT image. Right: PanSTARRS $z$ band image. The red plus symbol shows the location of the source.}
\label{fig:275.125}
\end{figure}

% Don't change these lines
\bsp	% typesetting comment
\label{lastpage}
\end{document}